\documentclass[aps,prd,superscriptaddress,onecolumn,floatfix,nofootinbib,preprintnumbers]{revtex4}

\usepackage{amsmath,graphicx,tabularx,url}

\begin{document}

\def\tablename{Table}
\def\figurename{Figure}

\def\gtot{\Gamma_\text{tot}}
\def\brinv{\text{BR}_\text{inv}}
\def\brsm{\text{BR}_\text{SM}}
\def\bratio{\mathcal{B}_\text{inv}}
\def\as{\alpha_s}
\def\az{\alpha_0}
\def\gz{g_0}
\def\w{\vec{w}}
\def\sdag{\Sigma^{\dag}}
\def\s{\Sigma}
\newcommand{\psib}{\overline{\psi}}
\newcommand{\Psib}{\overline{\Psi}}
\newcommand\one{\leavevmode\hbox{\small1\normalsize\kern-.33em1}}
\newcommand{\Mpl}{M_\mathrm{Pl}}
\newcommand{\p}{\partial}
\newcommand{\lag}{\mathcal{L}}
\newcommand{\qqquad}{\qquad \qquad}
\newcommand{\qqqquad}{\qquad \qquad \qquad}

\newcommand{\qb}{\bar{q}}
\newcommand{\matx}{|\mathcal{M}|^2}
\newcommand{\really}{\stackrel{!}{=}}
\newcommand{\msbar}{\overline{\text{MS}}}
\newcommand{\qns}{f_q^\text{NS}}
\newcommand{\lqcd}{\Lambda_\text{QCD}}
\newcommand{\met}{\slashchar{p}_T}
\newcommand{\pmiss}{\slashchar{\vec{p}}_T}

\newcommand{\st}[1]{\tilde{t}_{#1}}
\newcommand{\stb}[1]{\tilde{t}_{#1}^*}
\newcommand{\nz}[1]{\tilde{\chi}_{#1}^0}
\newcommand{\cp}[1]{\tilde{\chi}_{#1}^+}
\newcommand{\cm}[1]{\tilde{\chi}_{#1}^-}

% all the masses 
\providecommand{\mg}{m_{\tilde{g}}}
\providecommand{\mst}{m_{\tilde{t}}}
\newcommand{\msn}[1]{m_{\tilde{\nu}_{#1}}}
\newcommand{\mch}[1]{m_{\tilde{\chi}^+_{#1}}}
\newcommand{\mne}[1]{m_{\tilde{\chi}^0_{#1}}}
\newcommand{\msb}[1]{m_{\tilde{b}_{#1}}}

% units of measure
\newcommand{\mev}{{\ensuremath\rm MeV}}
\newcommand{\gev}{{\ensuremath\rm GeV}}
\newcommand{\tev}{{\ensuremath\rm TeV}}
\newcommand{\fb}{{\ensuremath\rm fb}}
\newcommand{\ab}{{\ensuremath\rm ab}}
\newcommand{\pb}{{\ensuremath\rm pb}}
\newcommand{\sign}{{\ensuremath\rm sign}}
\newcommand{\ifb}{{\ensuremath\rm fb^{-1}}}

% really great macro by Chris Lester
\def\slashchar#1{\setbox0=\hbox{$#1$}           % set a box for #1
   \dimen0=\wd0                                 % and get its size
   \setbox1=\hbox{/} \dimen1=\wd1               % get size of /
   \ifdim\dimen0>\dimen1                        % #1 is bigger
      \rlap{\hbox to \dimen0{\hfil/\hfil}}      % so center / in box
      #1                                        % and print #1
   \else                                        % / is bigger
      \rlap{\hbox to \dimen1{\hfil$#1$\hfil}}   % so center #1
      /                                         % and print /
   \fi}
\newcommand{\dslash}{\slashchar{\partial}}
\newcommand{\Dslash}{\slashchar{D}}

\def\eg{{\sl e.g.} \,}
\def\ie{{\sl i.e.} \,}
\def\etal{{\sl et al} \,}

\title{Boosted Semileptonic Tops in Stop Decays}

\author{Tilman Plehn}
\affiliation{Institut f\"ur Theoretische Physik, 
             Universit\"at Heidelberg, Germany}

\author{Michael Spannowsky}
\affiliation{Department of Physics and Institute of Theoretical Science, 
             University of Oregon, Eugene, USA}

\author{Michihisa Takeuchi}
\affiliation{Institut f\"ur Theoretische Physik, 
             Universit\"at Heidelberg, Germany}

%\date{\today}

\begin{abstract}
 Top partner searches are one of the key aspects of new physics
 analyses at the LHC. We correct an earlier statement that
 supersymmetric top searches based on decays to semileptonic tops are
 not promising. Reconstructing the direction of the boosted leptonic
 top quark and correlating it with the measured missing transverse
 energy vector allows us to reduce the top pair background to an
 easily manageable level. In addition, reconstructing the full
 momentum of the leptonic top quark determines the stop mass based on
 an $M_{T2}$ endpoint. 
\end{abstract}

\maketitle

%%%%%%%%%%%%%%%%%%%%%%%%%%%%%%%%%%%%%%%%%%%%%%%%%%%%%%%%%%%%%%%%%%%%%%%%%%%%%%%%

Searches for top squarks at hadron colliders aim at a fundamental
questions of electroweak symmetry breaking: if the Higgs boson
should be a fundamental scalar, how can its mass be stabilized? In
particular, is the Higgs mass protected by some symmetry? Such
symmetries typically predict the existence of a top partner, like in
supersymmetric or little Higgs models~\cite{review,meade}. In such a
case, studying the properties of top partners allows us to unravel the
nature of such an underlying fundamental symmetry protecting the
fundamental Higgs mass at the scale of electroweak symmetry
breaking.\bigskip

At the Tevatron, low-mass stop searches look for loop-induced stop
decays~\cite{stop_decays} to charm quarks and the lightest
neutralino~\cite{stops_tevatron_c}. Slightly larger stop masses more
in agreement with theoretical expectations makes it more promising to
look for decays to a bottom jet and the lightest
chargino~\cite{stops_tevatron_b}. For leptonic chargino decays this
signatures is irreducible from a leptonic top decay. Finally, if the
stop becomes heavier and the strong decay in a top quark and a gluino
is not yet kinematically allowed, the stop can decay into a top quark
and the lightest neutralino~\cite{stop_decays}. Because it includes
on-shell top quarks as stop decay products this channel is the hardest
to extract at the LHC. For hadronically decaying top quarks we know
that this final state has the advantage that we can fully reconstruct
its kinematics, which puts us into a promising position to study
angular correlations in the stop pair final
state~\cite{heptop}. Extending this analysis strategy to semileptonic
stop decays is the aim of this study.

There have been several suggestions as to what we might be able to say
about the nature of the stop based on a momentum reconstruction of its
visible decay products~\cite{stops_measure}; the first ingredient to
any such analysis is a way to extract stop pairs from the overwhelming
top pair background, the second the reconstruction of the stop
4-momentum. Over almost the entire supersymmetric parameter space
standard analysis methods fail at both tasks. On the other hand, fully
hadronic top pairs we can tag and fully reconstruct using fat jet
analysis tools~\cite{heptop,fatjet_top,fatjet_tth}.  Using the {\sc
  HEPTopTagger}~\cite{fatjet_tth,heptop} we can extract the stop
signature for a wide range of stop masses, reconstruct the top
4-momentum with high precision, and reconstruct the stop mass using a
simple $M_{T2}$ endpoint analysis~\cite{heptop}. Obviously, we can use
the same strategy for the hadronic side of signatures related to
semileptonic top pairs.\bigskip

The key idea behind the reconstruction of hadronically decaying heavy
particles using fat jets is to study the clustering history of a jet
algorithm for geometrically large jets including boosted heavy
particles. This idea was originally developed for $W$ bosons and top
quarks~\cite{fatjet1,fatjet_wash} and very successfully applied to Higgs decays to
bottoms~\cite{fatjet_higgs}, hadronic top
decays~\cite{fatjet_top,fatjet_tth,heptop} as well as supersymmetric
particle decays~\cite{fatjet_susy}. 

The obvious question is what we can learn from boosted hadronically
decaying top analyses for the equally boosted leptonic top
decay~\cite{tops_semilep}.  The main shortcoming of the traditional
$W$ and top mass constrained reconstruction is that we cannot use the
missing transverse momentum vector as a cut against backgrounds after
we use it for the reconstruction.  The aim is to instead limit
ourselves to the particular features of collinear leptonic top decays
to reconstruct the top
kinematics~\cite{thaler_wang,rehermann_tweedie}. 

In this paper we will proceed in two steps: first, we show how to
reconstruct the direction of the top 3-momentum exploiting the typical
kinematic features of boosted leptonic top decays. A cut on this top
direction serves as the main ingredient to the extraction of
semileptonic stop pairs. Second, again using typical features of
boosted top decays we reconstruct the full 4-momentum of the top
quark. In analogy to hadronically decaying tops this reconstruction
could be a starting point to many more phenomenological top pair
analyses.

%%%%%%%%%%%%%%%%%%%%%%%%%%%%%%%%%%%%%%%%%%%%%%%%%%%%%%%%%%%%%%%%%%%%%%%%%%%%%%%%
\section{Semileptonic stop decays}
\label{sec:intro}

To illustrate the power of reconstructing leptonic top pairs we show
how we can extract the stop signature
\begin{equation}
 pp \to \st1 \st1^*
    \to ( t \nz1) \, (\bar{t} \nz1) 
    \to ( b \ell^+ \nu \nz1) \, (\bar{b} j j \nz1) 
       +( b j j \nz1) \, (\bar{b} \ell^- \bar{\nu} \nz1) 
\end{equation}
including missing energy from top decay and from the dark matter agent
from top pair production. Our analysis will proceed based on two
strategies: first, we suggest a simple approximation of the top decay
kinematics to reconstruct the direction of the top 3-momentum
(Section~\ref{sec:dir}) and the full 4-momentum
(Section~\ref{sec:mom}). This way we arrive at a two-step strategy to
distinguish the stop signal from top pair backgrounds based on the
reconstructed top direction and on its reconstructed
4-momentum.\bigskip

Typical semileptonic analyses start from a set of acceptance cuts
requiring at least four jets and a charged lepton~\cite{heptop}. In
our case we include a hadronic top tag, which means we require
\begin{itemize}
\item[1.] exactly one trigger lepton \; ($p_T > 20$~GeV, $|\eta|<2.5$
  and isolation $E_T/E_{T,\ell} < 0.1$ within $R=0.2$)
\item[2.] missing transverse momentum $\met > 150$~GeV
\item[3.] one tagged hadronic top \; (mass drop,
  $p_T>200$~GeV, {\sc HEPTopTagger} default~\cite{heptop}, {\sc Fastjet}~\cite{fastjet})
\item[4.] one $b$ tag among the leading 3 jets outside the tagged top \; (C/A with
  $R=0.5$, $p_T>25$~GeV, $|\eta|<2.5$)
\item[5.] a bottom-lepton invariant mass allowed by the top decay
  kinematics $m_{b\ell}<\sqrt{m_t^2-m_W^2}=154.6$~GeV.
\end{itemize}
The missing energy cut in combination with the lepton requirement
safely removes all QCD backgrounds, including $b$ pair production. In
our analysis we do not take into account fakes from pure QCD events.

The main Standard Model background after these first analysis steps is
$W$+jets production, with an inclusive $W$ cross section roughly 200
times the top pair production rate. The probability to fake a tagged
top from $W$+jets events ranges around $4\%$, after requiring to
observe a fat jet with $p_T>200$~GeV~\cite{heptop}. The tagging
efficiency and mis-tagging rate for the {\sc HEPTopTagger} have to be
confirmed by a full experimental study and should in addition be
tested on a sufficiently large $t \bar{t}$ sample at the
LHC. Including the additional $b$ tag we expect semileptonic
$t\bar{t}$ pairs to become the main background to our stop searches.
The bottom jet isolation from the top fat jet is defined
geometrically. For the $b$ tag we assume a (mis-) tagging probability
of ($60\%,2\%$) as an event weight factor. Charm jets with a
generically higher mis-tagging probability should not spoil our
analysis because of their suppression compared to high-multiplicity
gluon production.\bigskip

%-----------------------------------------------------
\begin{table}[b]
\begin{tabular}{l|rrrr|rr|cc}
\hline 
& \multicolumn{4}{c|}{$\st1 \st1^*$}
& $t\bar{t}$ 
& $W$+jets 
& $S/B$ 
& $S/\sqrt{B}_{20\ifb}$ \\
\hline 
$\mst[\gev]$ & 340 & 440 & 540 & 640 &&& 440 & 440 \\
\hline
 0. cross section  &   5090&   1280&    402&    146& $9.2\cdot 10^5$ & $2.1 \cdot10^5$& 0.001 &$3.8$\cr
\hline
1. one lepton      & 1471&    373&    118&     42.5& $2.6\cdot 10^5$& $1.3\cdot 10^5$      & 0.001 &$2.7$\cr
2. $\met > 150$~GeV&     569&    239&     90.2&     35.5&   9825&   4512                          & $0.017$&$8.9$\cr
3. hadronic top tag&     74.5&     38.0&     16.8&      7.72&   1657&    141                                    &$0.021$&$4.0$\cr
4$^\prime$. tagged $b$ jet (100\%,0\%) &        47.2&     24.3&     11.3&      5.22&   1057&    0.00&$0.023$&$3.3$\cr
4. tagged $b$ jet (60\%,2\%) &    31.2&     15.9&      7.33&      3.38&    668&      4.35                          &$0.024$&$2.7$\cr
5. $m_{b\ell} < m_{b\ell}^{\max}$ &       27.5&     13.7&      6.34&      2.90&    642&      2.61                          &$0.021$&$2.4$\cr
\hline
\end{tabular}
\caption{Signal (for different stop masses) and backgrounds for the
  base analysis. All numbers given in fb.  For $W$+jets, we only show
  numbers after pre-selection ($n_\text{jets}\ge 3$).}
\label{tab:res1}
\end{table}
%-----------------------------------------------------

The status after these first five analysis steps we show in
Table~\ref{tab:res1}.  The stop with different masses we assume to
decay with a 100\% branching ratio to a top quark and a 98~GeV
lightest neutralino. The NLO normalization of the {\sc Herwig++}
~\cite{herwig} signal sample we obtain from {\sc
  Prospino}~\cite{prospino}. The top pair sample from {\sc
  Alpgen-Pythia} we normalize to the approximate NNLO rate around
918~pb~\cite{top_rate}. For subleading $W$+jets background we use the
default {\sc Alpgen-Pythia} normalization~\cite{alpgen}.  For signal
and background we include initial and final state radiation,
hadronization and underlying event. The top and stop samples we
generate inclusively without restricted decays.\bigskip

Table~\ref{tab:res1} shows how the main analysis steps act on the
signal and the dominant backgrounds. Indeed, we see that while the
lepton and missing energy requirements are easily passed by all
backgrounds, the hadronic top tag and the $b$ tag significantly reduce
the $W$+jets backgrounds and leave us with the dominant top pair
contamination. For the hadronic top tag we observe the expected large
efficiency for heavier stop samples as compared to continuum
$t\bar{t}$ production, which comes from the harder $p_{T,t}$ spectrum.

While these initial results are promising, Table~\ref{tab:res1}
clearly indicates that we definitely need another set of analysis
steps to improve the signal-to-background ratio to a manageable level,
to be able to cope with systematic experimental and theoretical
uncertainties.

%%%%%%%%%%%%%%%%%%%%%%%%%%%%%%%%%%%%%%%%%%%%%%%%%%%%%%%%%%%%%%%%%%%%%%%%%%%%%%%%
\section{Reconstructing the top direction}
\label{sec:dir}

The main difference between the reconstruction of a hadronically
decaying top and a leptonic top decays is the significantly smaller
number of observables. Aside from the boost of the top and its
correlations with other particles in the final state we have to rely
on three measurements
\begin{equation}
E_\ell \qqqquad  E_b \qqqquad m_{b\ell} \; ,
\label{eq:measurements}
\end{equation}
assuming that we know the bottom and lepton 4-vectors after
identifying the particles. The invariant mass can be replaced by the
angle between the lepton and $b$ momenta.

There have been a few attempts to find strategies to identify boosted
tops decaying leptonically. One of them are `stuck leptons', referring
to leptons which are not well isolated from the $b$ jet from the top
decay~\cite{thaler_wang,rehermann_tweedie}.  The invariant mass of the
visible top decay products $m_{b \ell}$ is not necessarily close to
its upper bound if the neutrino is not observed. Nevertheless, the
mass drop $m_b^2/m^2_{b\ell}$ ranges around zero for boosted tops and
$W$+jets while it tends towards one for heavy flavor jets. Because
Eq.(\ref{eq:measurements}) tells us that there exist three distinctive
kinematic observables we can combine the mass drop measurement with
preferably sizeable $\Delta R_{b \ell}$ and $E_\ell/(E_\ell+E_b) \sim
0.4$ to construct a leptonic top tagger, but with a limited
yield~\cite{thaler_wang}. One strategy to improve the
tagging/rejection efficiencies is to include the tracker-level
mini-isolation observable $p_{T,\ell}/\sum p_{T, \text{charged}} >
0.9$ summed over all charged tracks within a cone around the lepton or
muon~\cite{rehermann_tweedie}.\bigskip

Our strategy is similar to those in the sense that we attempt to
reconstruct (parts of) the top 4-momentum without making use of the
measured missing transverse energy. This allows us to contrast the top
4-momentum with the measured missing energy vector as a background
rejection cut. However, we follow a different path to extract
information on the top 4-momentum. To describe the 3-momenta of the
top decay products we define the orthonormal coordinate system
\begin{alignat}{5}
\hat{p}^\text{D}
&= \frac{\vec{p}_{b \ell}}{|\vec{p}_{b\ell}|} 
&&\text{leading $\vec{p}_{b\ell}$ direction in $b-\ell$ decay plane} \notag \\
\hat{p}^\parallel
&= \frac{\vec{p}_{\ell} - (\vec{p}_{\ell} \cdot  \hat{p}^\text{D}) \, \hat{p}^\text{D}}
{|\vec{p}_{\ell} - (\vec{p}_{\ell} \cdot  \hat{p}^\text{D}) \, \hat{p}^\text{D}|}
\qquad 
&&\text{subleading direction in $b-\ell$ decay plane} \notag \\
\hat{p}^\perp
&= \hat{p}^\text{D} \times \hat{p}^\parallel
&&\text{subleading direction to $b-\ell$ decay plane.} 
\label{eq:coord}
\end{alignat}
Within this coordinate system the neutrino momentum is parameterized as
\begin{equation}
\vec{p}_\nu = x_\text{D} \hat{p}^\text{D} 
            + x_\parallel \hat{p}^\parallel
            + x_\perp \hat{p}^\perp. 
%            = p_\nu 
%            (\hat{x}_\text{D} \hat{p}^\text{D} 
%            + \hat{x}_\parallel \hat{p}^\parallel
%            + \hat{x}_\perp \hat{p}^\perp) \; .
\end{equation}
In the appendix we show the relative size of these three
components.\bigskip

To reconstruct the neutrino momentum, assuming we know the neutrino
mass, we need three components. Because we want to compare the
neutrino momentum from the reconstructed top to the measured
two-dimensional missing transverse momentum $\pmiss$ these are three
actual unknowns. The $W$ and top mass constraints give us two, so we
need to make at least one assumption to describe the collinear decay
of a boosted top. As shown in the appendix, we can choose two well
motivated approximations
\begin{alignat}{5}
%\vec{p}_\nu^\perp &= 0 \qquad \qquad 
x_\perp &= 0 \qquad \qquad 
&& \text{decay plane approximation} \notag \\
%\vec{p}_\nu^\parallel &= 0 
x_\parallel &= 0
&& \text{orthogonal approximation,} 
\label{eq:approx}
\end{alignat}
and correspondingly estimate the neutrino momentum as
\begin{equation}
\vec{p}_\nu^\text{est} \Big|_\parallel 
\qqquad \text{or} \qquad
\vec{p}_\nu^\text{est} \Big|_\perp \; .
\end{equation}
Based on this estimated neutrino momentum we can construct
measurements which distinguish for example the stop signal from the
top pair background where the main difference is the existence of
additional sources of missing momentum in the signal.\bigskip

Before constructing such observables we need to sketch a few features
of the two approximations for the neutrino and top momentum
estimates. For example, the mass constraints are quadratic
constraints, so we cannot expect a unique solution for the neutrino
and top momenta. Instead, there will be a set of four discrete
solutions from which we need to extract the physical solution.  The
decay plane approximation, which as we will discuss in the appendix is
motivated by the Jacobian peak in the top momentum we consistently
accept the solution which returns the smallest top momentum.  For the
orthogonal approximation, which rejects the $t\bar{t}$ background most
efficiently, we take the solution which gives the smaller subtracted
missing momentum $\met - p_{T,\nu}^\text{est}$.\bigskip

Reconstructing the direction of the neutrino 3-momentum or the top
3-momentum is of course not a full reconstruction as we know it from
the hadronic top decays using the {\sc HEPTopTagger}. However, even
with its reduced numbers of observables it is sufficient for example
to reconstruct the azimuthal angle between the estimated neutrino or
top direction and the measured missing transverse energy
\begin{equation}
\Delta \phi \equiv \Delta \phi (\pmiss, \hat{p}^\text{est}) 
=  \phi(\pmiss) - \phi(\hat{p}^\text{est}) \; .
\label{eq:del_phi}
\end{equation}
The estimated momentum entering this azimuthal difference is not
uniquely fixed. We can use the estimated neutrino momentum, the
estimated leptonically decaying top momentum, or the momentum of the
lepton-bottom system.

\begin{figure}[t]
\includegraphics[width=0.30\textwidth]{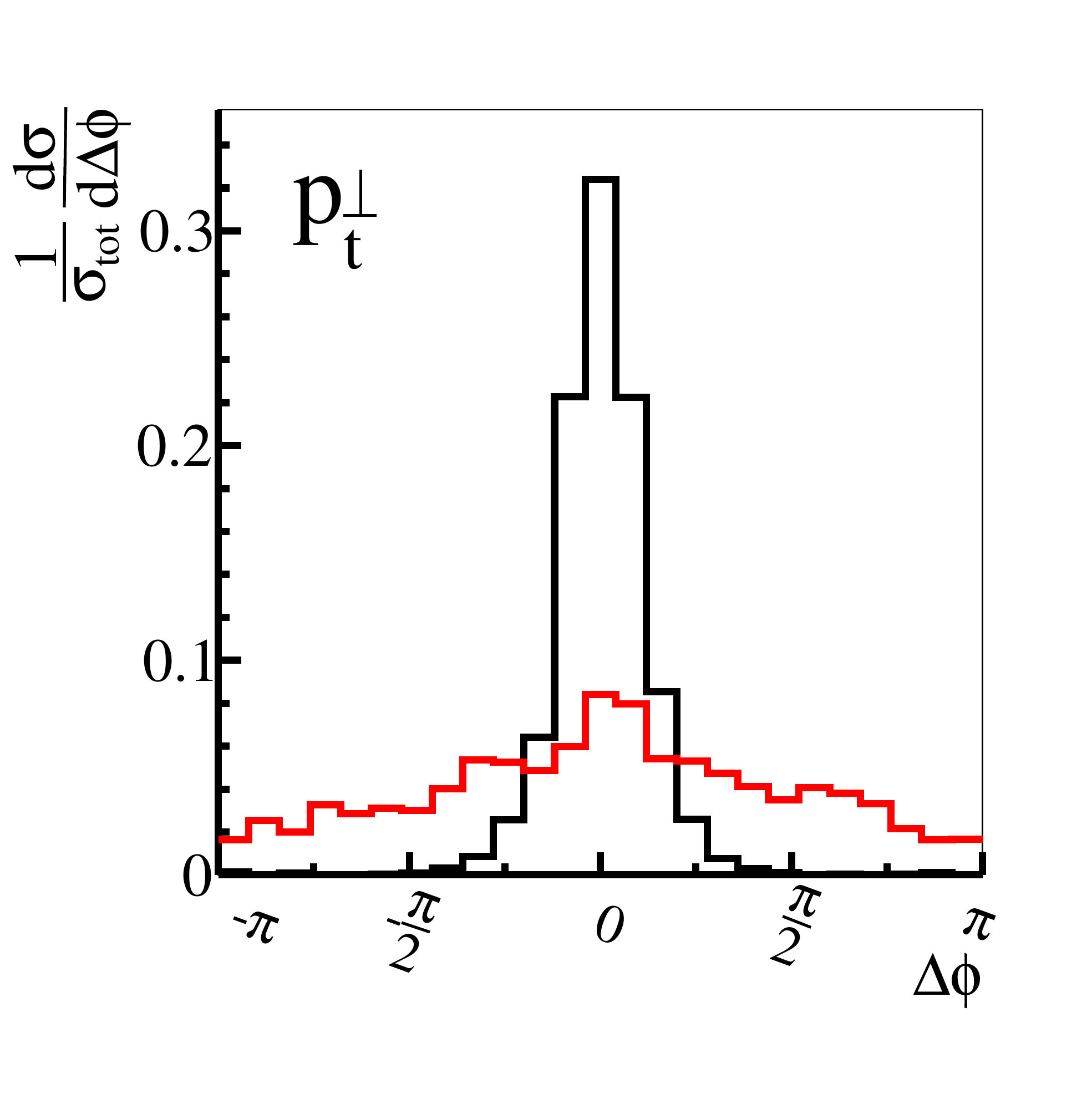}
\hspace*{0.025\textwidth}
\includegraphics[width=0.30\textwidth]{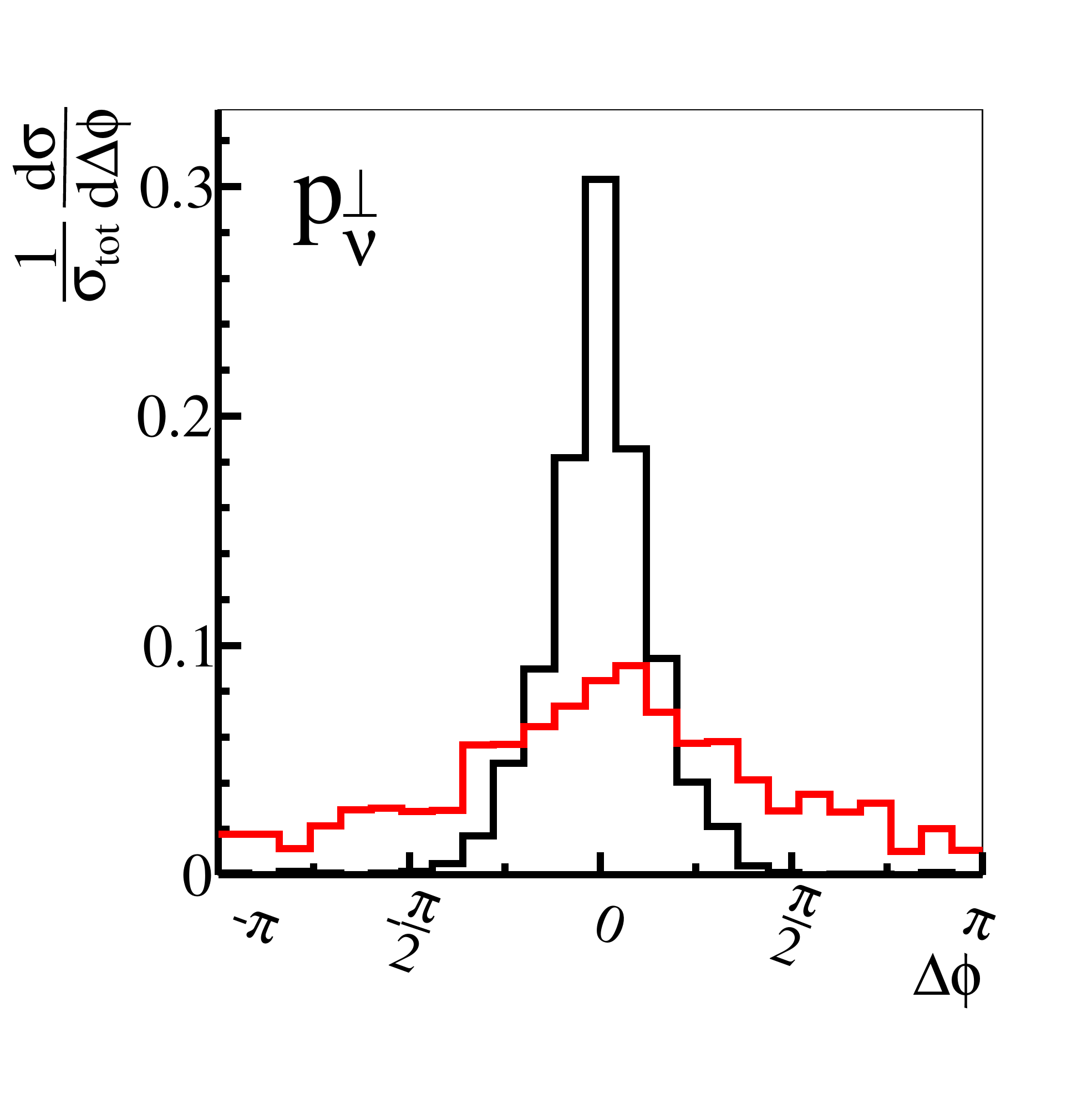}
\hspace*{0.025\textwidth}
\includegraphics[width=0.30\textwidth]{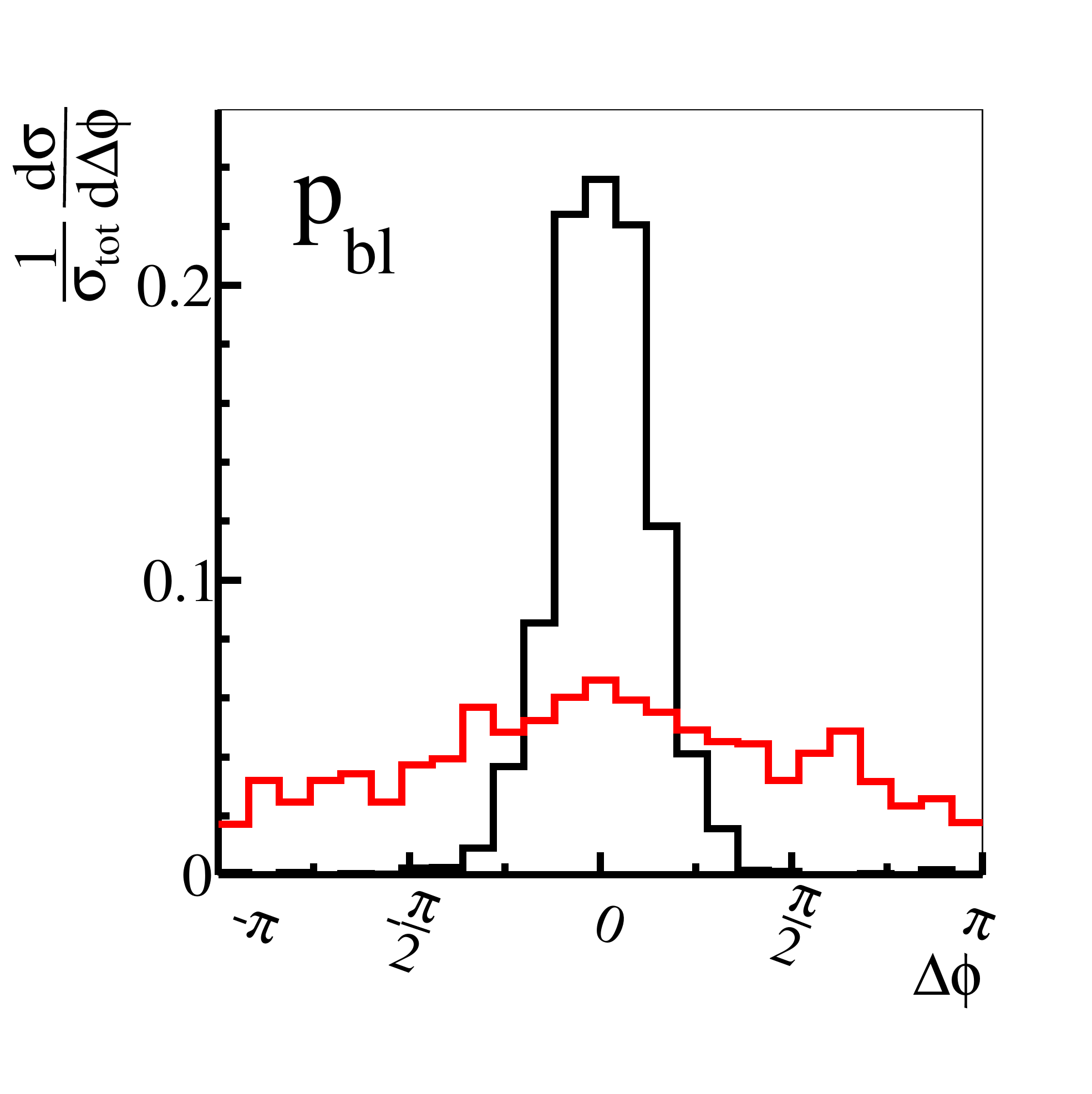} \\[-3mm]
\includegraphics[width=0.30\textwidth]{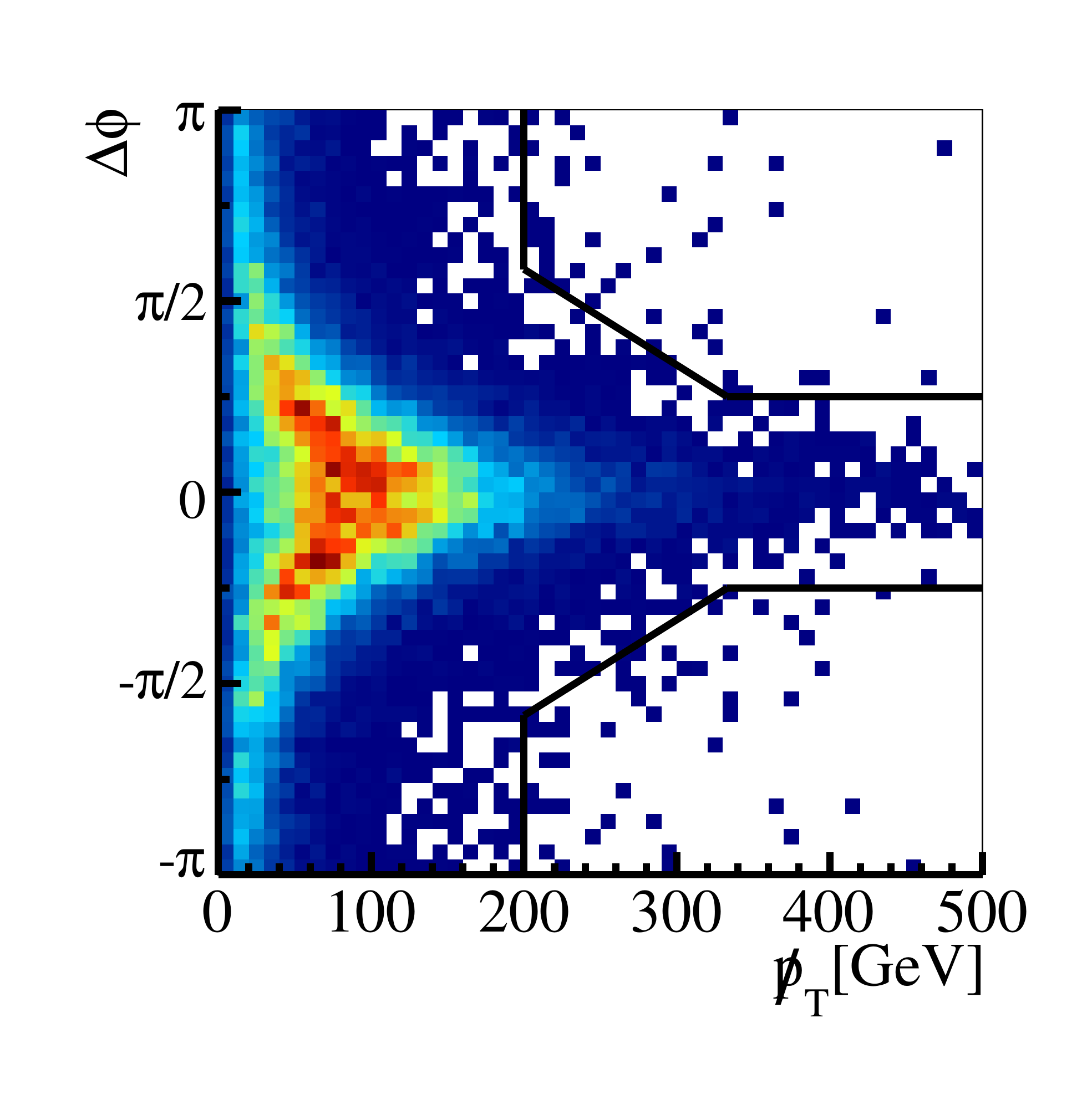}
\hspace*{0.025\textwidth}
\includegraphics[width=0.30\textwidth]{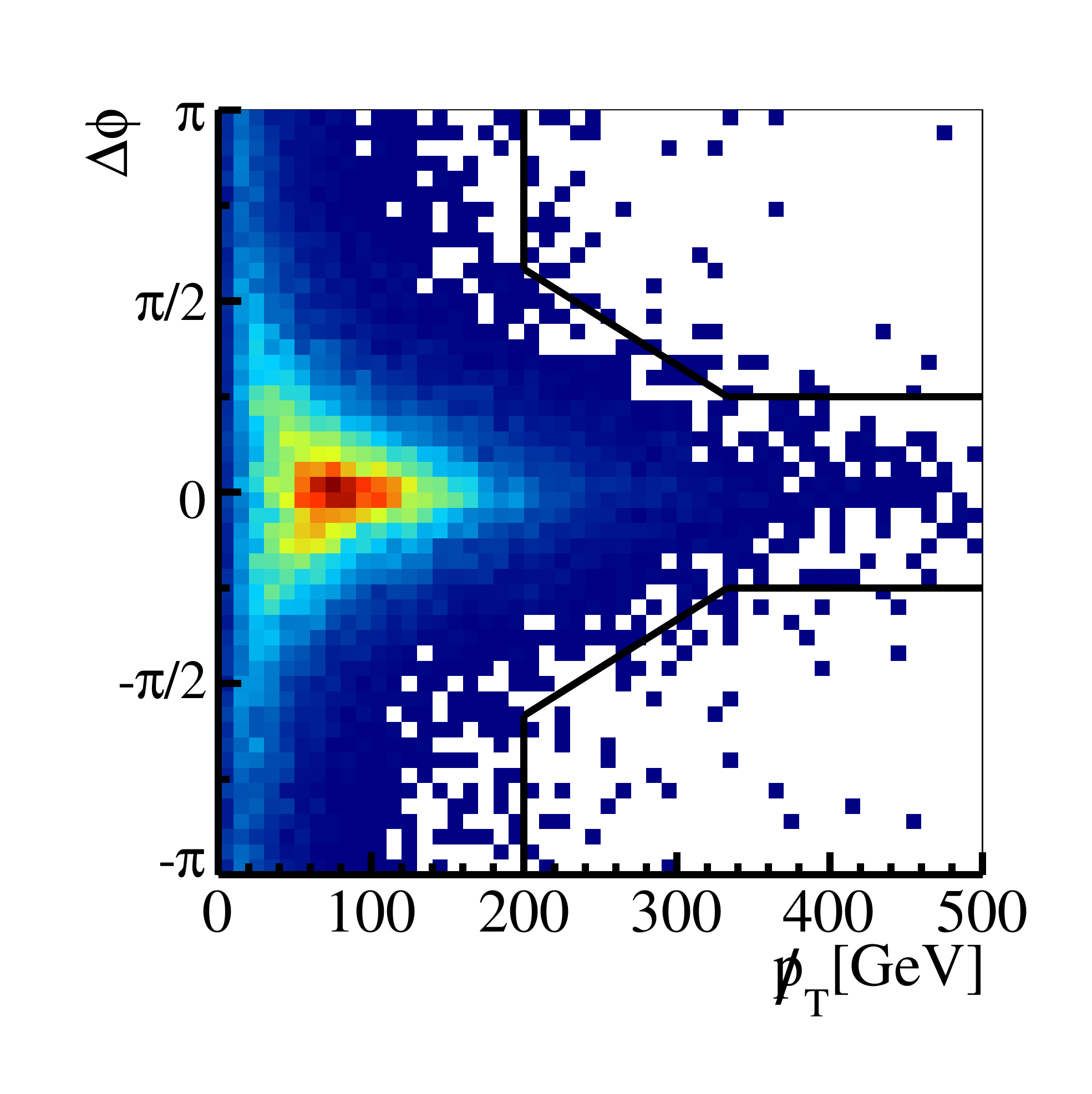}
\hspace*{0.025\textwidth}
\includegraphics[width=0.30\textwidth]{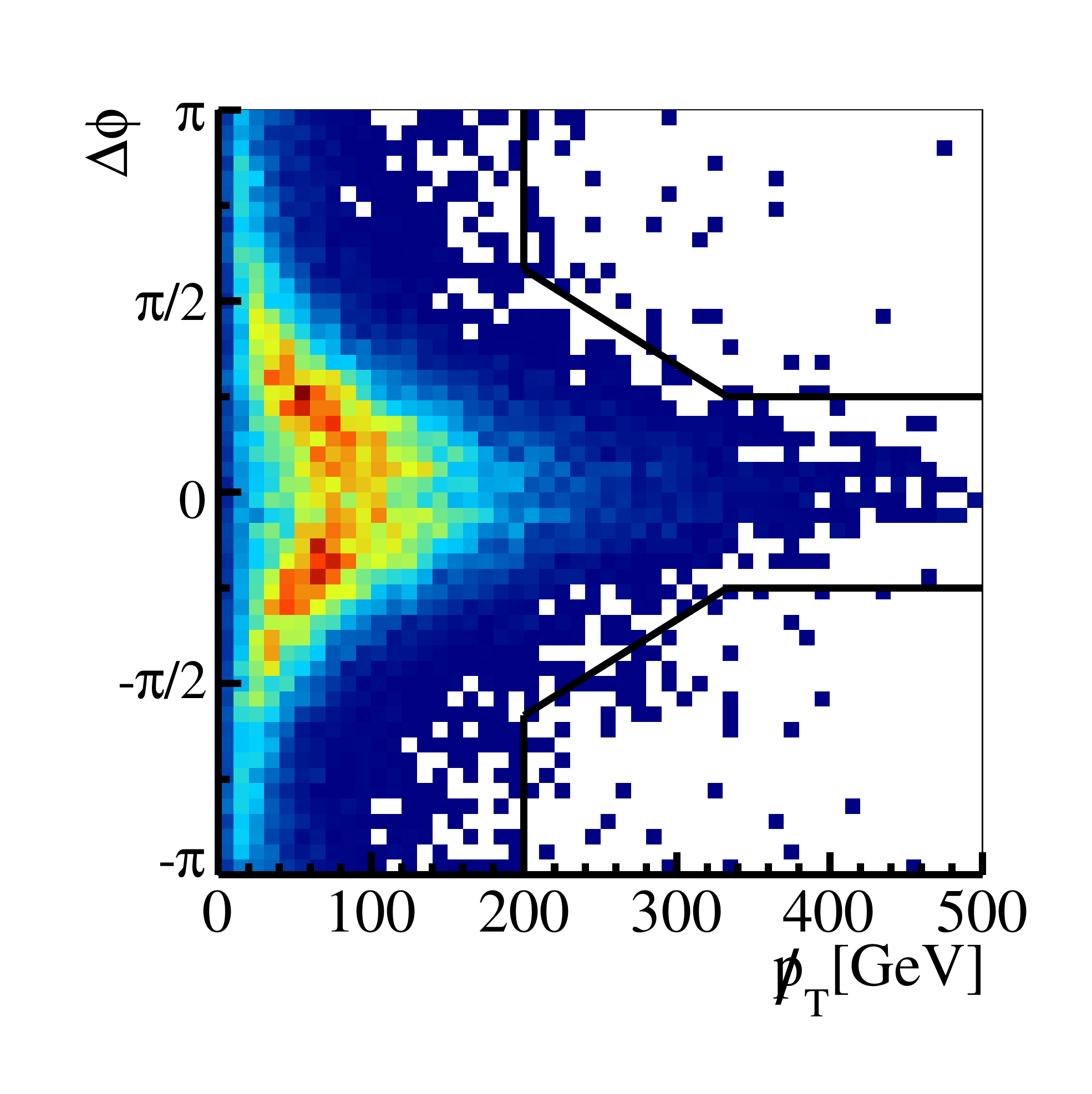} \\[-3mm]
\includegraphics[width=0.30\textwidth]{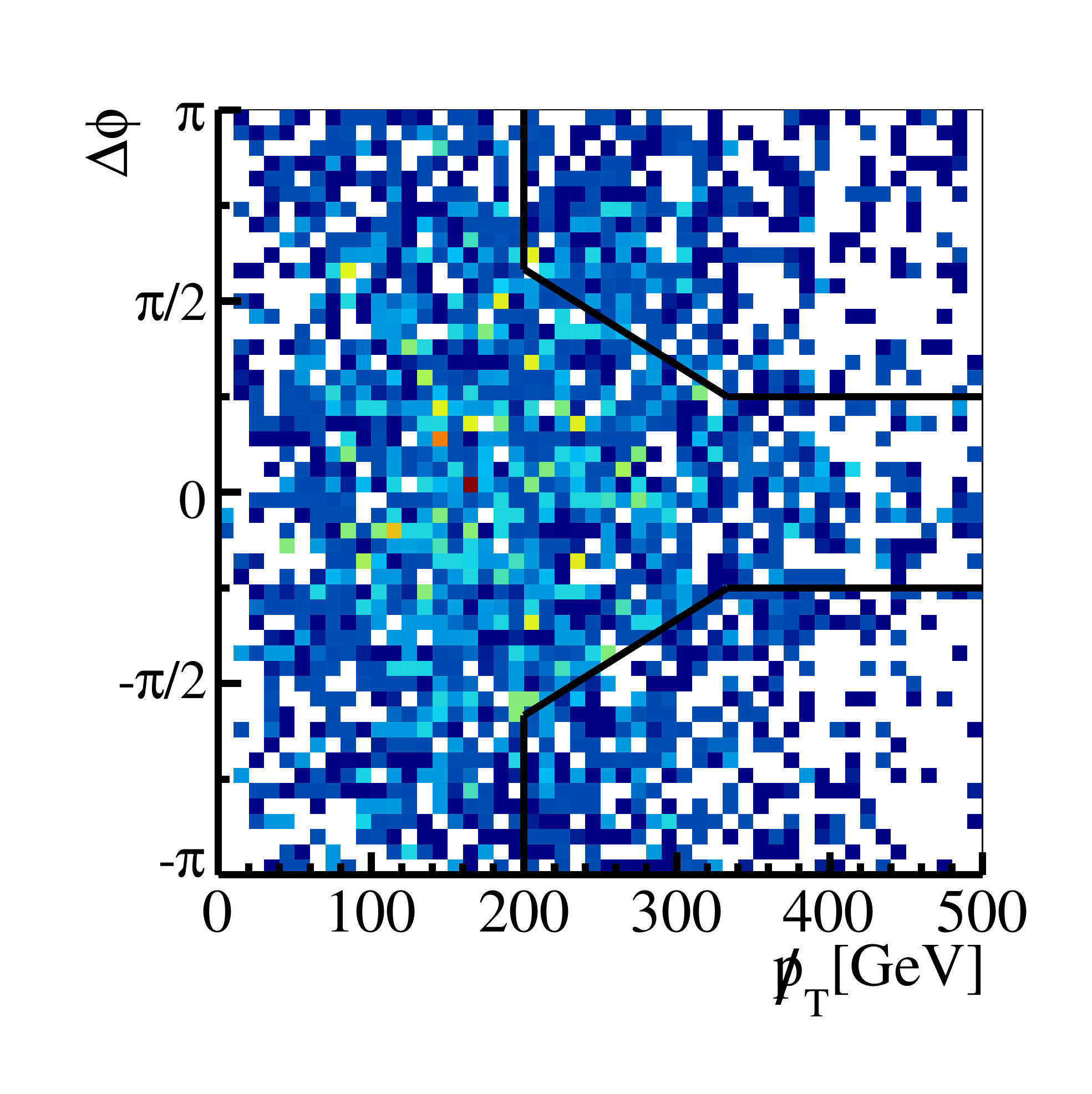}
\hspace*{0.025\textwidth}
\includegraphics[width=0.30\textwidth]{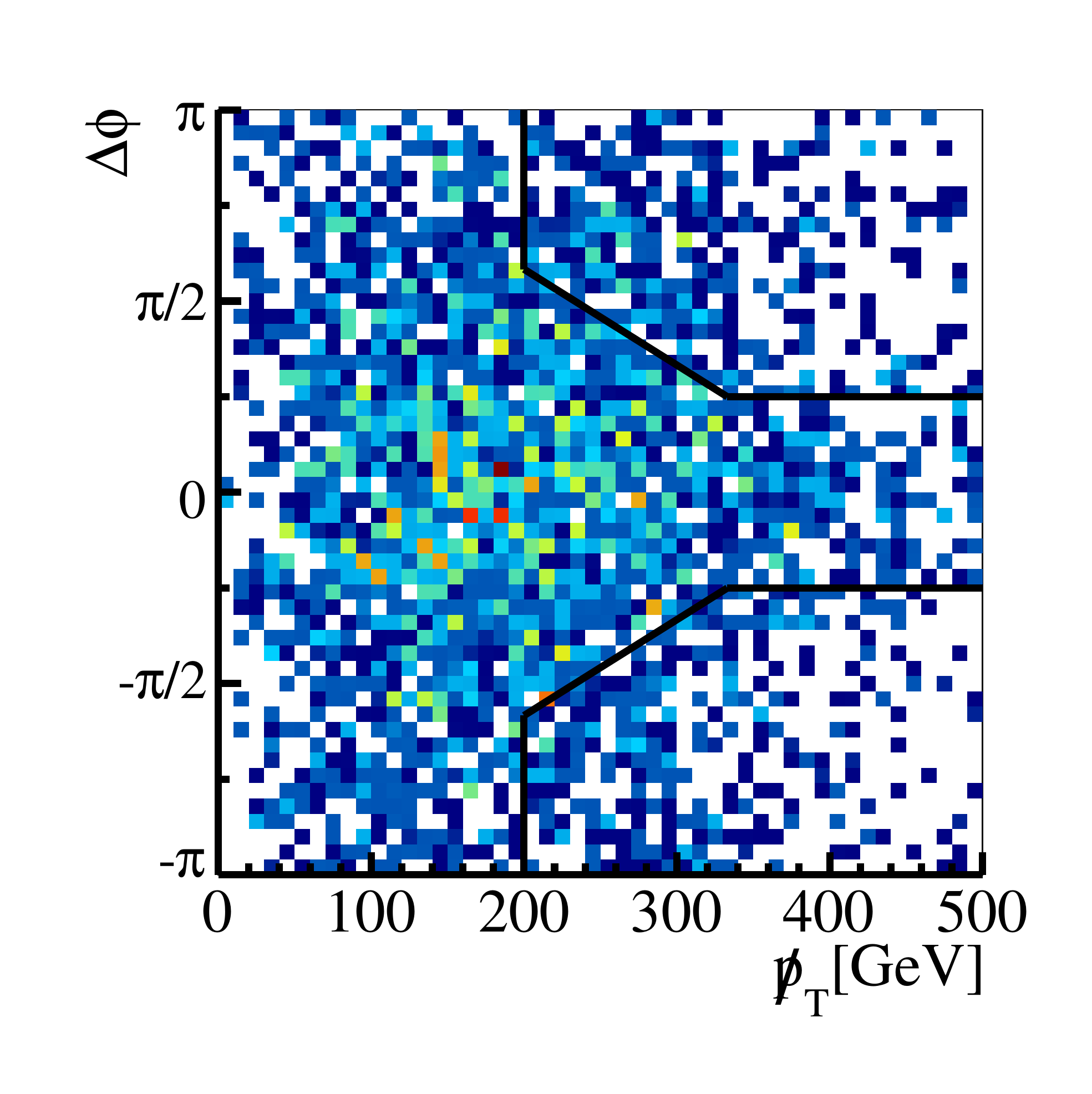}
\hspace*{0.025\textwidth}
\includegraphics[width=0.30\textwidth]{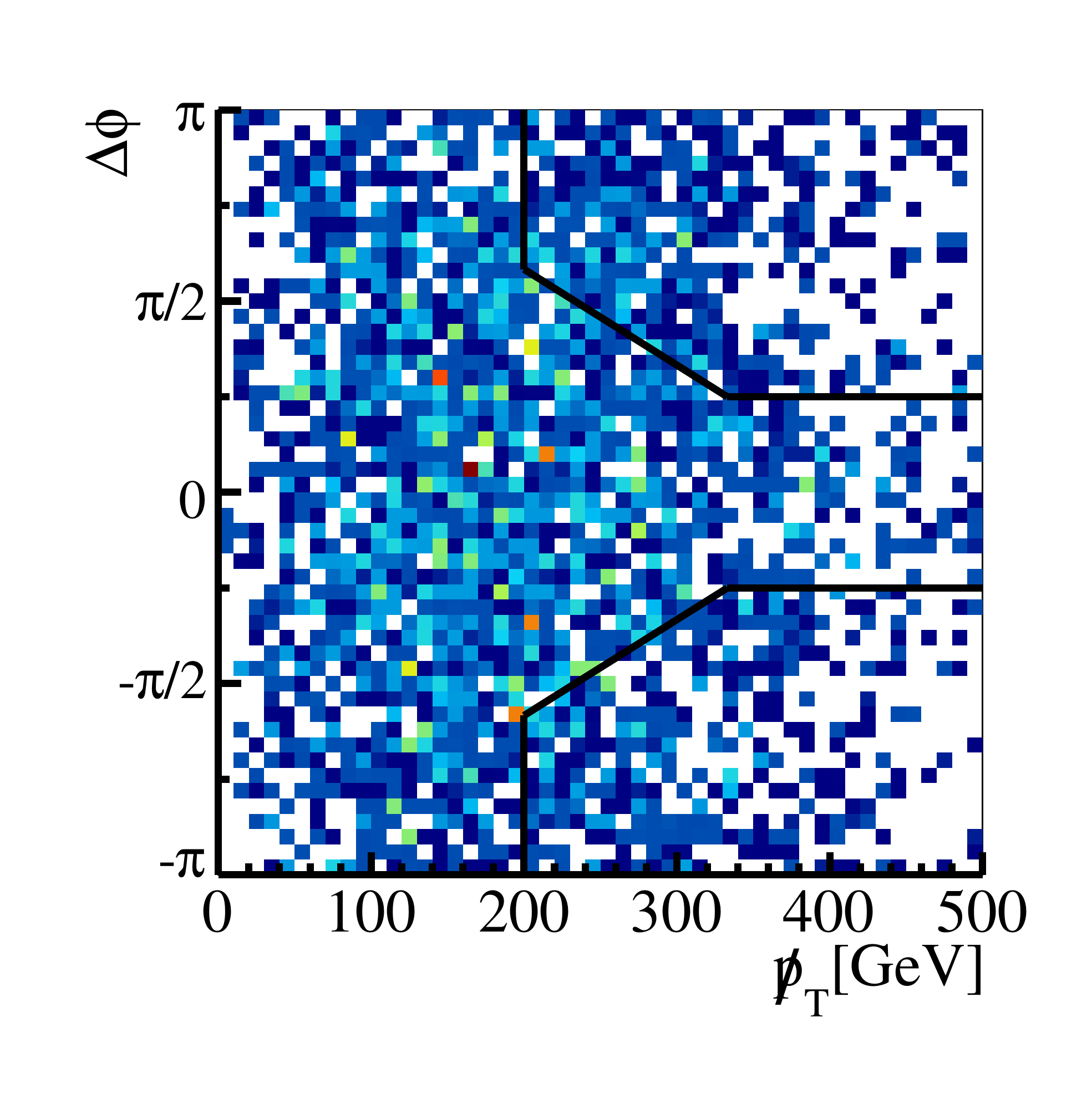}
\vspace*{-5mm}
\caption{Top: one-dimensional $\Delta\phi$ distributions with an
  additional cut $\met > 200$~GeV for the stop signal (red) and the
  top pair background (black). Below: two-dimensional $\met$
  vs. $\Delta\phi$ distributions for the $t\bar{t}$ background (2nd
  row) and for the $\tilde{t}\tilde{t}^\ast$ signal (3rd row).  From
  left to right we show $\Delta \phi(\pmiss,\hat{p}^{\perp}_{t,\ell})$, $\Delta
  \phi(\pmiss,\hat{p}^{\perp}_\nu)$ and $\Delta \phi(\pmiss,\hat{p}_{b\ell})$. 
   All plots are
  at the hadron level and based on the orthogonal approximation.  We
  apply the universal selection cuts 1-5 and the border of the cut 8 is shown in the two-demensional plots.}
\label{fig:metphi}
\end{figure}

From the discussion in the appendix estimating the top direction
appears to be more promising than estimating the neutrino direction,
simply because the neutrino momentum with its large relative
uncertainty constitutes only a small part of the top momentum. In
addition, the studies in the appendix suggest to use the orthogonal
approximation for this azimuthal angle study. As a starting point we
look at the correlations between the measured missing transverse
energy vector and the reconstructed leptonic top directions. It turns
out that this observable has the best yield for the stop pair search.
In the top row of Figure~\ref{fig:metphi} we show the $\Delta \phi$
distributions for the stop signal and the top backgrounds for the
orthogonal approximation defined in Eq.(\ref{eq:approx}). It shows a
clear difference between signal and $t\bar{t}$ events for all three
choices. The signal shows hardly a peak because the reconstructed
momentum does not correspond to a physical momentum and is hence
uncorrelated.  The width of the background peak which we need to cut
out is smallest when we include the estimated top direction, as
compared to the neutrino or bottom-lepton directions. This includes a
hadron level simulation, but the picture might of course change once
we include a detector simulation. Qualitatively, all three
reconstructed directions work at a similar level.\bigskip

Improving over this one-dimensional $\Delta \phi$ distribution we can
include the correlation with the measured missing transverse momentum.
It exploits the following feature: for semileptonic top pair
events, if the originating top is not boosted enough, there exists an
upper bound for the neutrino momentum.  To observe large $\met$, the
originating top momentum has to be aligned with the $\met$ direction.
Again, in Figure~\ref{fig:metphi} we show the results for three
reconstructed momentum directions, which work qualitatively the
same.\bigskip

%-----------------------------------------------------
%-----------------------------------------------------

These observations we can turn into an analysis strategy to extract
semileptonic stop pairs after applying the five generic steps shown in
Table~\ref{tab:res1}. First of all, it is not guaranteed that given a
measured bottom and lepton momentum configuration we find a valid
reconstructed neutrino momentum. As a consistency requirement we
should check if the reconstructed neutrino momentum combined with the
observed bottom and lepton momentum really corresponds to the kind of
boosted top quarks which we assume to exist in the first place:
\begin{itemize}
\item[6.] possible solution for respective approximation. 
\item[7.] $p_{T,t}^\text{est} > 200$~GeV. 
\end{itemize}
The orthogonal and decay plane approximations perform very differently
under the first of these two requirements. By construction and as
explained in the appendix, the signal efficiency of the orthogonal
approximation ranges around 50\% while the decay plane approximation
almost always delivers a valid solution for the neutrino
momentum. However, from Table~\ref{tab:res2} we see that this larger
signal efficiency is at least compensated by the also larger
background efficiency, so that in the final result the orthogonal
approximation delivers a better signal-to-background ratio as well as
a better significance for an integrated luminosity of $20~\ifb$.  The
situation might well be different for smaller luminosities where the
number of signal events after all cuts becomes the main
hurdle.\bigskip

As the last step of our analysis, following Figure~\ref{fig:metphi} we
require a two-dimensional condition for the azimuthal separation
defined in Eq.(\ref{eq:del_phi}) correlated with the measured missing
transverse momentum:
\begin{itemize}
\item[8.] one tagged leptonic top fulfilling all three conditions
\begin{equation*}
|\Delta \phi|> \frac{13}{12}\pi-\frac{\met}{400~\text{GeV}}\pi 
\qquad \qquad \qquad 
\met>200~\gev 
\qquad \qquad \qquad 
|\Delta \phi|>\frac{\pi}{4}.
\end{equation*}
\end{itemize}
All signal and background cross sections after these signal specific
cuts we show in Table~\ref{tab:res2}. As long as we are only
interested in the direction of the leptonic top momentum but not its
size the orthogonal approximation works better, which is also shown in
Figure~\ref{fig:dphi} in the appendix. Table~\ref{tab:res3} shows that
over a wide range of stop masses we find of the order of 20 signal
events for an integrated luminosity of $20~\ifb$ at a collider energy
of 14~TeV. The signal-to-background ratio from the orthogonal
approximation ranges around $S/B \sim 1-2$ and the Gaussian
significance exceeds $S/\sqrt{B}\sim 5$ for stop masses from 340 to
540~GeV. For larger stop masses the kinematics for the signal as
compared to the softer continuum top pair production becomes more
distinctive, so our analysis should be adapted to this parameter
regime. Note that the numbers in Table~\ref{tab:res3} are computed at
the hadron level. Applying all cuts 1.-8. gets rid of essentially all
semileptonic top pair events at parton level, so the events shown as
$t\bar{t}$ background in Table~\ref{tab:res3} are mostly mis-measured
leptonic top pairs.

Also in Table~\ref{tab:res3} we see how the decay plane approximation
performs slightly worse for the signal-to-background ratio, but gives
a similar significance based on a slightly larger number of signal
events after all cuts.  While this significance for semileptonic stop
decays might be slightly smaller than for the purely hadronic
mode~\cite{heptop} the signal-to-background ratio is better and the
efficiency of the critical hadronic top tagger is reduced to a linear
dependence. Combining the two analysis strategies means that based on
the order of $10~\ifb$ of LHC data at 14~TeV (or an adjusted number
for a lower collider energy) we can discover a top squark mainly
decaying to top quarks and a dark matter agent in both of the two
dominant top decay channels. This not only would give us faith in such
an observation, it is also the starting point for detailed studies of
such a new physics signal.

%-----------------------------------------------------
\begin{table}[t]
\begin{tabular}{l|rrrr|rr|r||rrrr|rr|r}
\hline 
& \multicolumn{7}{c||}{orthogonal approximation}
& \multicolumn{7}{c}{decay plane approximation} \\
\hline 
& \multicolumn{4}{c|}{$\st1 \st1^*$}
& $t\bar{t}$ 
& $W$+jets 
& $S/B$
& \multicolumn{4}{c|}{$\st1 \st1^*$}
& $t\bar{t}$ 
& $W$+jets 
& $S/B$ \\
\hline 
$\mst[\gev]$ & 340 & 440 & 540 & 640 &&& 440 
             & 340 & 440 & 540 & 640 &&& 440 \\
\hline
1.-5. base cuts &     27.38&     13.71&      6.33&      2.89&    642.72&      2.63&$0.021$&&&&&&\cr
6. approximation        &     14.81&     7.69&      3.61&      1.66&    285.16&      1.41&$0.027$
&     27.33&     13.67&      6.31&      2.89&    642.37&      2.63&0.021\cr
7. ${p^\text{est}_T>200}$GeV     &   8.61&      4.53&      2.41&      1.24&    215.62&      0.60&$0.021$
&     9.13 &      5.16&      2.87&      1.61 &    242.21&      0.54&0.021\cr
8. $\met$ vs. $\Delta\phi$ cut   
& 0.97&      1.52&      1.23&      0.76&      0.72&      0.02 &2.06
& 1.22 & 1.82 & 1.53 & 1.02 & 1.31 & 0.06 & 1.33
\\
\hline
\end{tabular}
\caption{Signal (for different stop masses) and backgrounds after the
  leptonic top tag. We show both approximations to the top momentum
  defined in Eq.(\ref{eq:approx}).  The $\met$ vs. $\Delta \phi$ cut
  is defined in the text.  All numbers given in fb.}
\label{tab:res2}
\end{table}
%-----------------------------------------------------

%-----------------------------------------------------
\begin{table}[b]
\begin{tabular}{l|cccc||cccc}
\hline 
& \multicolumn{4}{c||}{orthogonal approximation}
& \multicolumn{4}{c}{decay plane approximation} \\
\hline 
$\mst[\gev]$ & \; 340 \; & \; 440 \; & \; 540 \; & \; 640 \; 
             & \; 340 \; & \; 440 \; & \; 540 \; & \; 640 \; \\
\hline
$S_{20\ifb}$  
&19.4        &   30.4	&24.6 &15.2
&24.4	&    36.4	&30.6 &20.4\\
$S/B$  
&1.31 &	2.05	& 1.66 &1.03
&0.89 &	1.33	&1.12 &0.74\\   
$S/\sqrt{B}_{20\ifb}$
&5.04&	7.90	& 6.34 &3.95
& 4.66 &	6.95	& 5.84 &3.90\\
\hline
\end{tabular}
\caption{Results based on the reconstructed top momentum direction for
  the two approximations to the neutrino momentum,
  Eq.(\ref{eq:approx}). All eight cuts shown in Tables~\ref{tab:res1}
  and~\ref{tab:res2} are included. The LHC energy is 14~TeV.}
\label{tab:res3} 
\end{table}
%-----------------------------------------------------

%%%%%%%%%%%%%%%%%%%%%%%%%%%%%%%%%%%%%%%%%%%%%%%%%%%%%%%%%%%%%%%%%%%%%%%%%%%%%%%%
\section{Reconstructing the top 4-momentum}
\label{sec:mom}

The strategy described in Section~\ref{sec:dir} we can also use to
fully reconstruct the 4-momentum of the leptonically decaying top
quark. An additional question becomes how much this additional
observable helps in extracting the stop signature, as compared to the
results shown in Table~\ref{tab:res3}. Up to now the downside to our
stop pair analysis is that we do not have a handle on the stop
mass. 

If as an additional step we want to gain more information on the heavy
particle decaying into a boosted top the $M_{T,2}$ endpoint is the
prime observable and requires full knowledge of the top
4-momentum~\cite{mt2}. As for the purely hadronic stop decay the
crucial observation is that this stransverse mass endpoint to work we
need a one-step stop decay into the invisible dark matter agent. A
pair of top squarks decaying into tagged tops and two dark matter
particles is precisely of this topology.\bigskip

For the endpoint measurement, we best rely on a combination of the
orthogonal and decay plane approximations. First, we use the
orthogonal approximation for the basic cuts 1.-5. and the more
specific top rejection cuts 6.-8. Then, we use either the orthogonal
or the the decay plane approximation to compute the $m_{T2}$
distribution, \ie we define a $m_{T2}$ variable in terms of the
estimated top momenta and the estimated missing momentum from other
sources than the neutrino as
\begin{equation}
M_{T2} \Big|_{\parallel,\perp} = 
M_{T2}(p_{t,h}^\text{HEPTop}, p_{t,\ell}^\text{est},\met - p_{T, \nu}^\text{est}) \Big|_{\parallel, \perp} \; .
\end{equation}
Once we measure the leptonic and hadronic top there should appear the
endpoint
\begin{equation}
M_{T2} < m_{\tilde t}\left(1 -\frac{m_{\tilde \chi^0_1}^2}{m_{\tilde t}^2}\right)
\end{equation}
in the stop signal events.  In Figure~\ref{fig:mt2} we show the
$M_{T2}$ signal and background resolutions in the orthogonal
approximation and in the decay plane approximation and clearly see the
signal endpoint around 418~\gev.  From a purely signal point of view
and without the two-dimensions $\Delta \phi$ vs. $\met$ cut (8.) the
endpoint of the stransverse mass defined in the decay plane
approximation is more precise.  The reason is that in the decay plane
approximation the estimated top momentum is by construction almost
always smaller than the true value while in the orthogonal
approximation, the estimated top momentum frequently exceeds the true
value, which spoils the endpoint.  Note that the signal-to-background
ratio $S/B$ is still fixed by the numbers for the orthogonal
approximation which we use for the event selection. This signal
feature survives after we include the final $\Delta \phi$ vs. $\met$
top rejection cut.

%-----------------------------------------------------
\begin{figure}[t]
\includegraphics[width=0.24\textwidth]{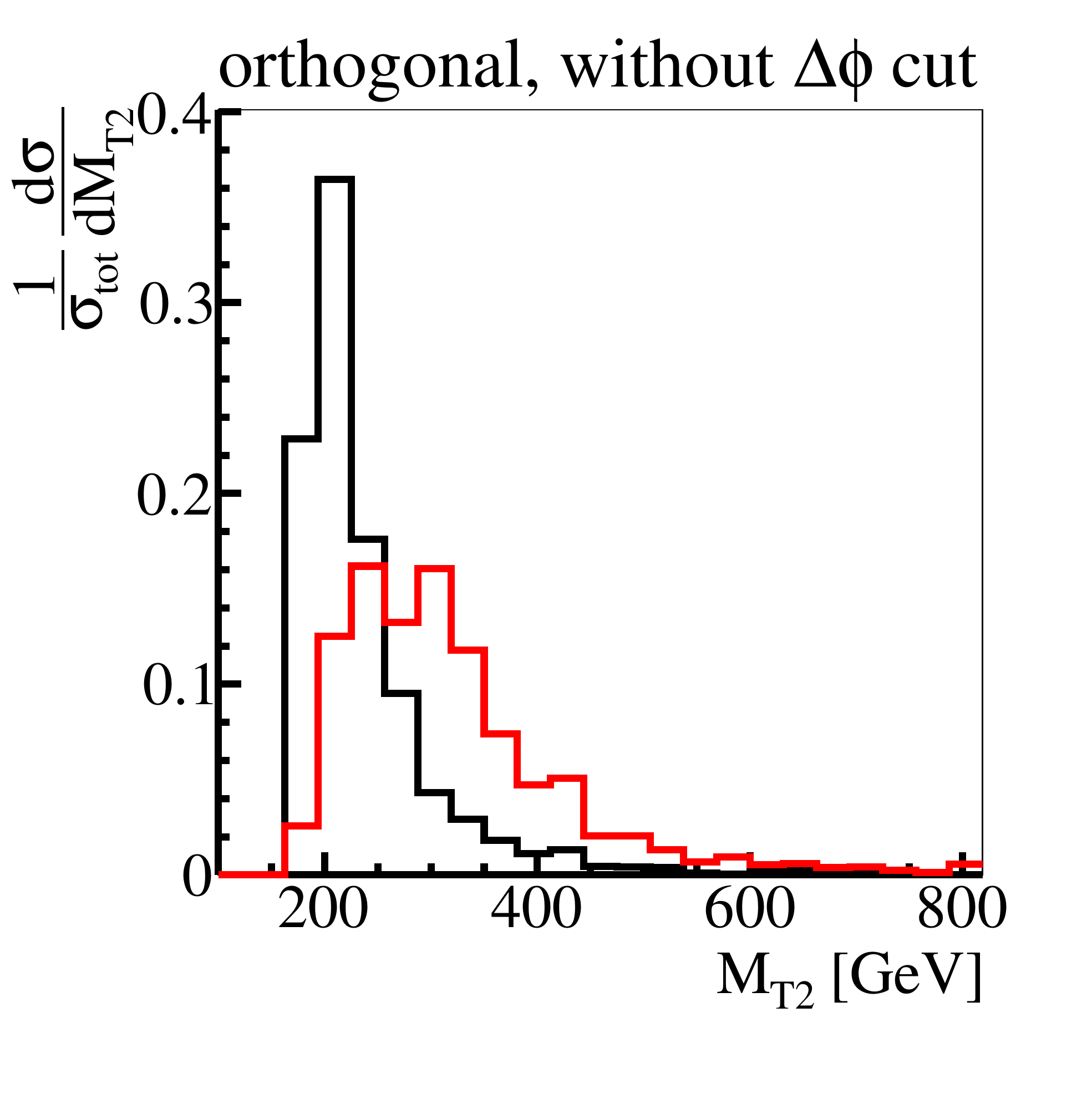}
\includegraphics[width=0.24\textwidth]{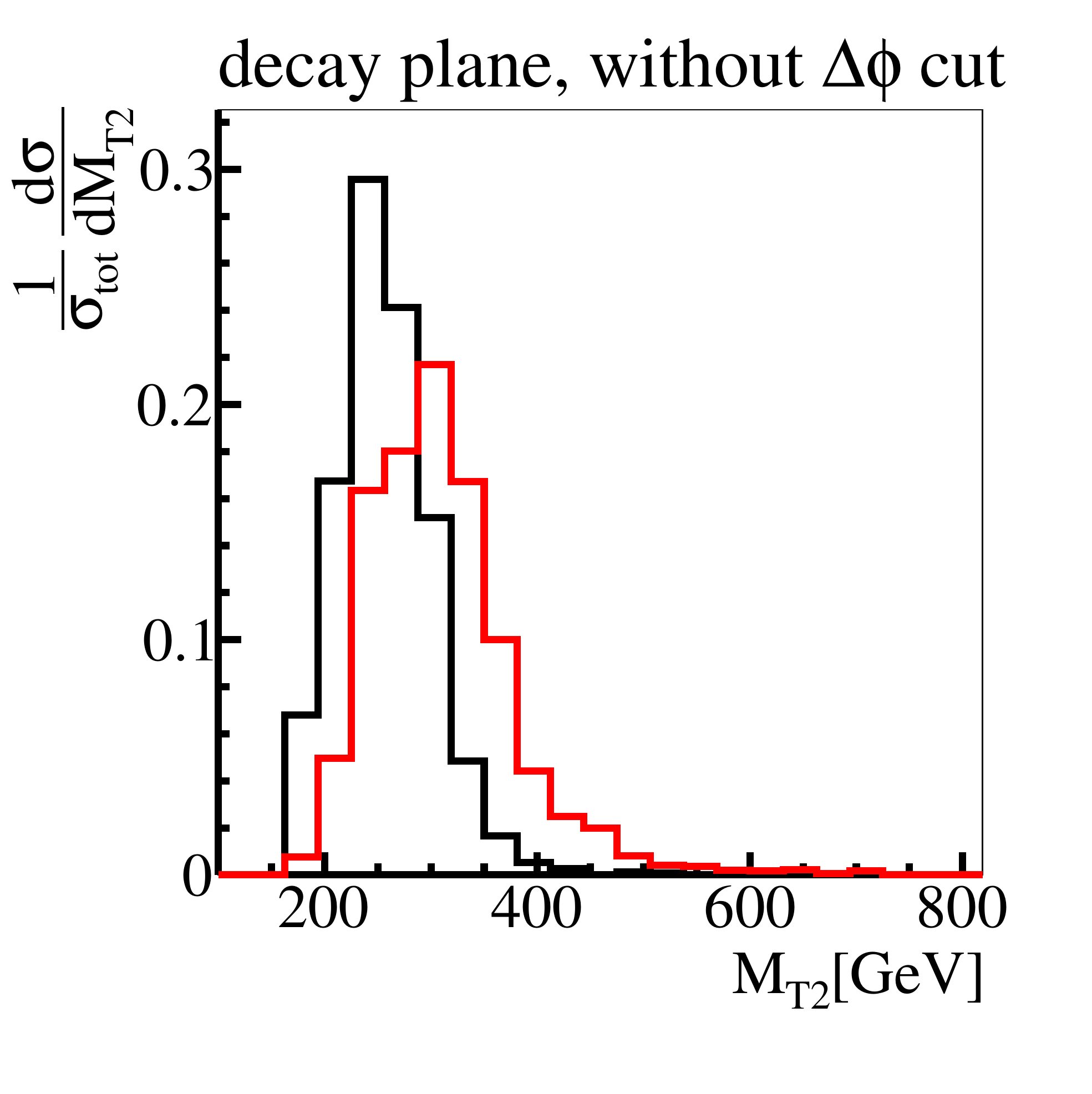}
\includegraphics[width=0.24\textwidth]{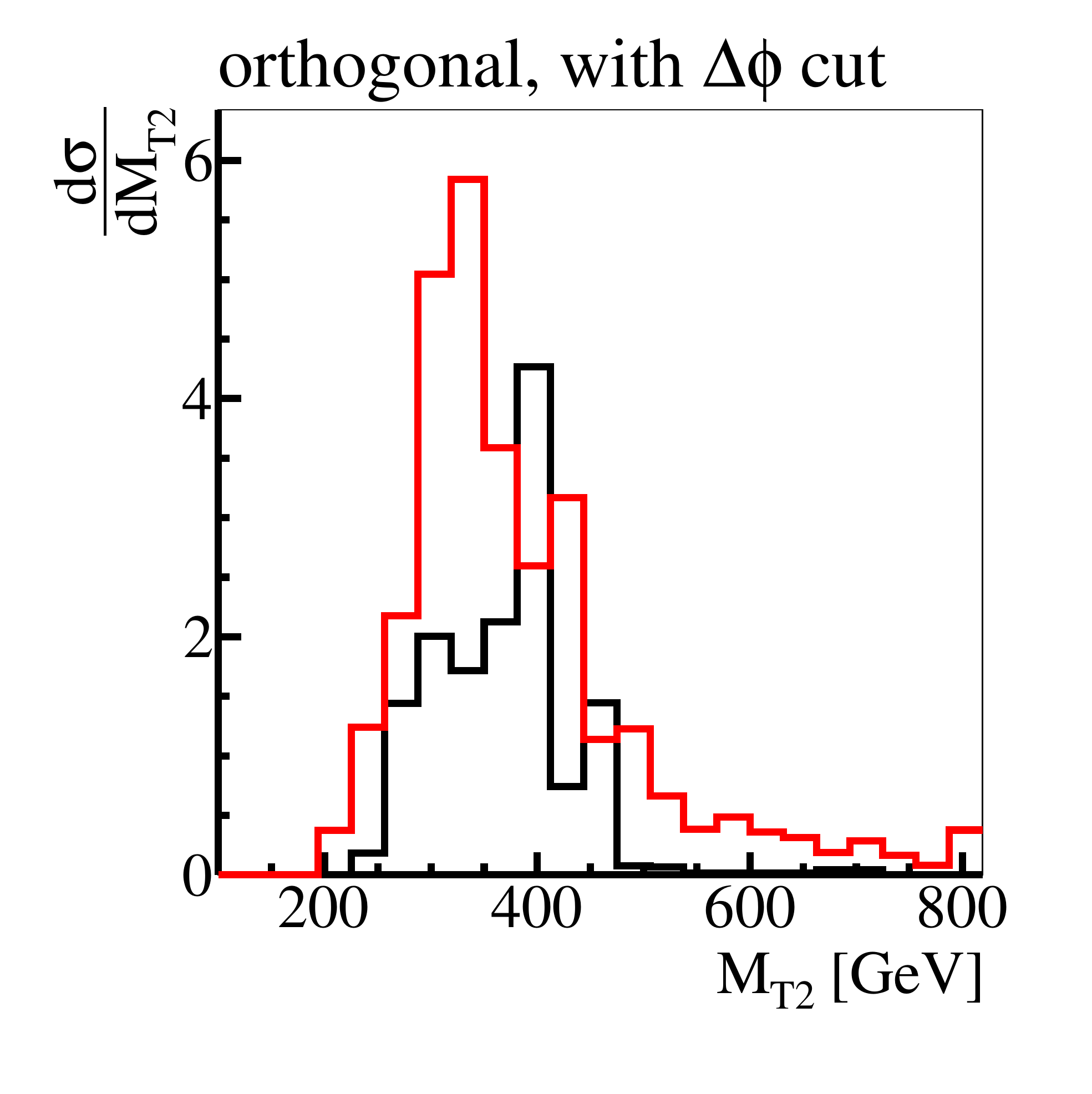}
\includegraphics[width=0.24\textwidth]{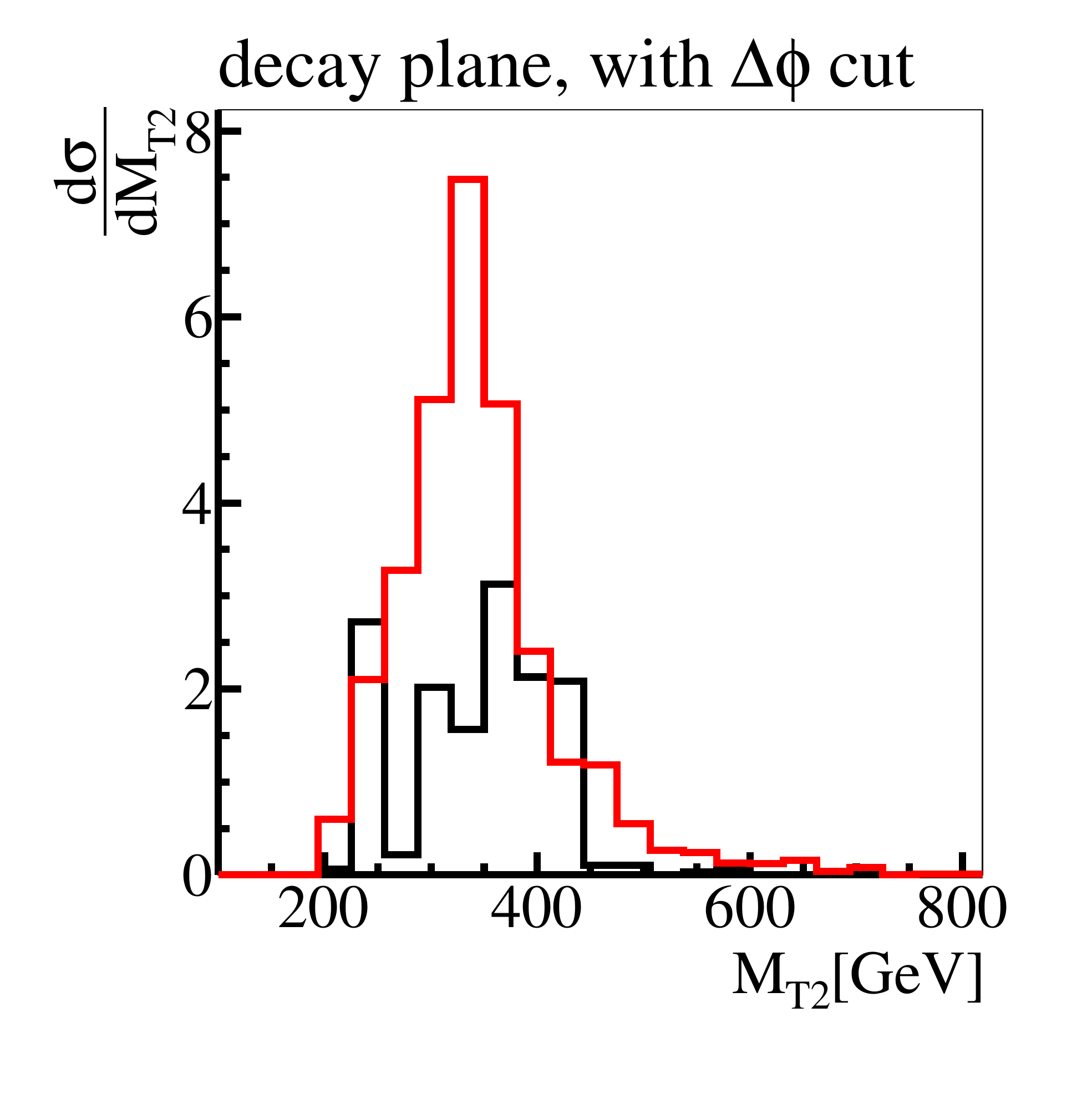}
\vspace*{-5mm}
\caption{$M_{T2}$ distributions for the stop signal (red) and the
  $t\bar{t}$ background (black).  From the left we show the orthogonal
  and the decay plane approximations without the $\Delta \phi$ vs
  $\met$ cut but with $\met > 200$~GeV and then both of them with the
  $\Delta \phi$ vs $\met$ cut. The expected endpoint is $M_{T2} <
  m_{\tilde t}(1 -{m_{\tilde \chi^0_1}^2}/{m_{\tilde t}^2})= 418~\gev$.}
\label{fig:mt2}
\end{figure}
%-----------------------------------------------------

%-----------------------------------------------------
\begin{table}[t]
\begin{tabular}{l|rrrr|rr|cc}
\hline 
& \multicolumn{4}{c|}{$\st1 \st1^*$}
& $t\bar{t}$ 
& $W$+jets 
& $S/B$ 
& $S/\sqrt{B}_{20\ifb}$ \\
\hline 
$\mst[\gev]$ & 340 & 440 & 540 & 640 &&& 440 & 440\\
\hline
1.-8. base and top rejection cuts       & 0.97&      1.52&      1.23&      0.76&      0.72&      0.02 &2.06 & 7.9\\
9. $M_{T2}|_{\parallel}>250$~GeV          &      0.65&      1.41&      1.20&      0.74&      0.64&      0.02   &2.13& 7.8  \cr
\phantom{9.} $M_{T2}|_{\parallel}>350$~GeV&      0.11&      0.59&      0.80&      0.58&      0.39&      0.02& 1.44 & 4.1\cr
\hline
\end{tabular}
\caption{Signal (for different stop masses) and backgrounds for the
  full stop pair analysis. All numbers given in fb.}
\label{tab:res4}
\end{table}
%-----------------------------------------------------

The top pair background at hadron level but without the
two-dimensional $\Delta \phi$ vs. $\met$ cut is slightly harder for
the decay plane approximation than for the orthogonal
approximation. Again, this feature generally survives the application
of the $\Delta \phi$ vs. $\met$ cut, but with a big caveat arising
from the experimental resolution which should dominate the behavior of
this large-$M_{T2}$ tail. From the right two panels of
Figure~~\ref{fig:mt2} we expect that the decay plane approximation
should work slightly better after applying all base cuts and top
rejection cuts, but the final word on this comparison needs to include
realistic detector effects. At this stage we rely on the orthogonal
approximation to determine the direction of the top, but on the decay
plane approximation for the magnitude of the top momentum.

As the final step of our stop pair analysis we need to check if the
$M_{T2}$ endpoint is useful for the actual signal extraction, \ie if
it improves either the significance or the signal-to-background
ratio. The results of such a cut we show in Table~\ref{tab:res4}.  We
see that the azimuthal separation of the top momentum from the missing
energy vector alone is highly efficient and gets hardly improved by
the additional $M_{T2}$ cut. The only reason to consider an $M_{T2}$
cut in the extraction of the stop signal would be to replace for
example the azimuthal direction cut, which means replacing the
leptonic top reconstruction of the azimuthal angle by the complete top
reconstruction by our tagging algorithm. Again, this might be an
option after taking into account all detector effects, but at the
level of our analysis the reconstruction of the azimuthal top
direction is more efficient.

%%%%%%%%%%%%%%%%%%%%%%%%%%%%%%%%%%%%%%%%%%%%%%%%%%%%%%%%%%%%%%%%%%%%%%%%%%%%%%%%
\section{Outlook}
\label{sec:outlook}

In this study we have corrected an earlier statement that at the LHC
it is not possible to observe semileptonic stop pairs over a wide
range of stop masses.  In the spirit of the hadronic stop pair
analysis~\cite{heptop} we develop a search strategy for stop pairs
decaying into a boosted semileptonic top pair plus missing energy. We
assume 14~TeV center of mass energy and of the order of $20~\ifb$ of
integrated luminosity. Different scenarios in collider energy and
luminosity can easily be explored.\bigskip

First of all, the hadronic top decay we fully reconstruct using the
standard {\sc HEPTopTagger}.  Earlier studies of boosted leptonic top
quarks~\cite{thaler_wang,rehermann_tweedie} have shown that when it
comes to reconstructing the full top 4-momentum the leptonic decay is
not too promising. Our approach therefore first focuses on a more
limited measurement, namely the azimuthal direction of the leptonic
top quark. This directional vector we can reconstruct based on a
decomposition of the neutrino momentum into a leading boost direction
and two orthogonal sub-leading directions. We show a detailed
comparison between two different approximations and come to the
conclusion that it is promising to reduce the number of kinematic
unknowns in the top decay by fixing the neutrino momentum orthogonally
to the lepton-bottom decay plane.

Based on this strategy we extract the stop pair signal from the
dominant irreducible top pair background with $5\sigma$ significance
for $20~\ifb$ of LHC data with a signal-to-background ratio above
unity for stop masses from 340 to 540~GeV. Our analysis is not
optimized for any stop mass, so this result should be very generic and
should be improved in particular for large stop masses. The discovery
potential for semileptonic stop is of the same order as for purely
hadronic stop decays, both of them based on boosted top quark
reconstruction.\bigskip

A full reconstruction of the top 4-momentum is considerably harder. On
the other hand, to measure the stop mass for example using the
classical $m_{T2}$ endpoint we develop such a strategy. While this
additional measurement has little impact on the significance or the
signal-to-background ratio with which we can extract the stop signal
it should allow us to determine the stop mass from an $m_{T2}$
endpoint based on a combination of the orthogonal and decay plane
approximation of the leptonic top decay.

The details of the stop mass measurement and its uncertainties need to
be determined by an appropriate detector simulation.  The endpoint
value will give us a stop mass measurement in combination with all
other supersymmetric mass measurements at the LHC. As mentioned in the
introduction, the stop mass is particularly useful when we combine it
with the Higgs mass because the two probe a similar set of model
parameters~\cite{sfitter}.\bigskip

The software used in this study will be published as part of the {\sc
  HEPTopTagger} analysis tool~\cite{heptop}.

%%%%%%%%%%%%%%%%%%%%%%%%%%%%%%%%%%%%%%%%%%%%%%%%%%%%%%%%%%%%%%%%%%%%%%%%%%%%%%%%
\newpage
\appendix

\section{Tagging leptonic tops}
\label{app:details}

In analogy to Ref.~\cite{heptop} we illustrate in detail the
motivation of the leptonic top tagger, its implementation, and its
performance. All results shown in this appendix are based on a hadron
level analysis, but without any detector simulation, just as the
analysis presented in the main body of this paper. In this appendix we
focus on the reconstruction of semileptonic top pairs, \ie we only
apply the leptonic top tagger to a sample of top pair events.\bigskip

As mentioned in the main text the $W$ and top mass conditions can be
written as
\begin{alignat}{5}
(p_\ell+p_b+p_\nu)^2=&m_t^2 \notag \\
(p_\ell+p_\nu)^2=&m_W^2 \; .
\label{eq:masscondition}
\end{alignat}
All three masses of the decay particles are known, which means that
once we measure the lepton and $b$ 4-momenta the allowed neutrino
3-momentum lies on the line which we will show can be described as the
intersection of an elliptic surface and a paraboloid.\bigskip

Following Eq.(\ref{eq:coord}), repeated here for convenience with
$\vec{p}_{b \ell} = \vec{p}_b + \vec{p}_\ell$,
\begin{alignat}{5}
\hat{p}^\text{D}
&= \frac{\vec{p}_{b \ell}}{|\vec{p}_{b\ell}|} 
&&\text{leading $\vec{p}_{b\ell}$ direction in $b-\ell$ decay plane} \notag \\
\hat{p}^\parallel
&= \frac{\vec{p}_{\ell} - (\vec{p}_{\ell} \cdot  \hat{p}^\text{D}) \, \hat{p}^\text{D}}
{|\vec{p}_{\ell} - (\vec{p}_{\ell} \cdot  \hat{p}^\text{D}) \, \hat{p}^\text{D}|}
\qquad 
&&\text{subleading direction in $b-\ell$ decay plane} \notag \\
\hat{p}^\perp
&= \hat{p}^\text{D} \times \hat{p}^\parallel
&&\text{subleading direction to $b-\ell$ decay plane.} 
\label{eq:coord2}
\end{alignat}
we can write for example the neutrino momentum as
\begin{equation}
\vec{p}_\nu = x_\text{D} \hat{p}^\text{D} 
            + x_\parallel \hat{p}^\parallel
            + x_\perp \hat{p}^\perp 
            = p_\nu 
            (\hat{x}_\text{D} \hat{p}^\text{D} 
            + \hat{x}_\parallel \hat{p}^\parallel
            + \hat{x}_\perp \hat{p}^\perp) \; .
\label{eq:def_xhat}
\end{equation}
In Figure~\ref{fig:x2x3} we show the two subleading $\hat{x}$
distributions for different in the originating top momentum.\bigskip 

%-----------------------------------------------------
\begin{figure}[t]
\includegraphics[width=0.24\textwidth]{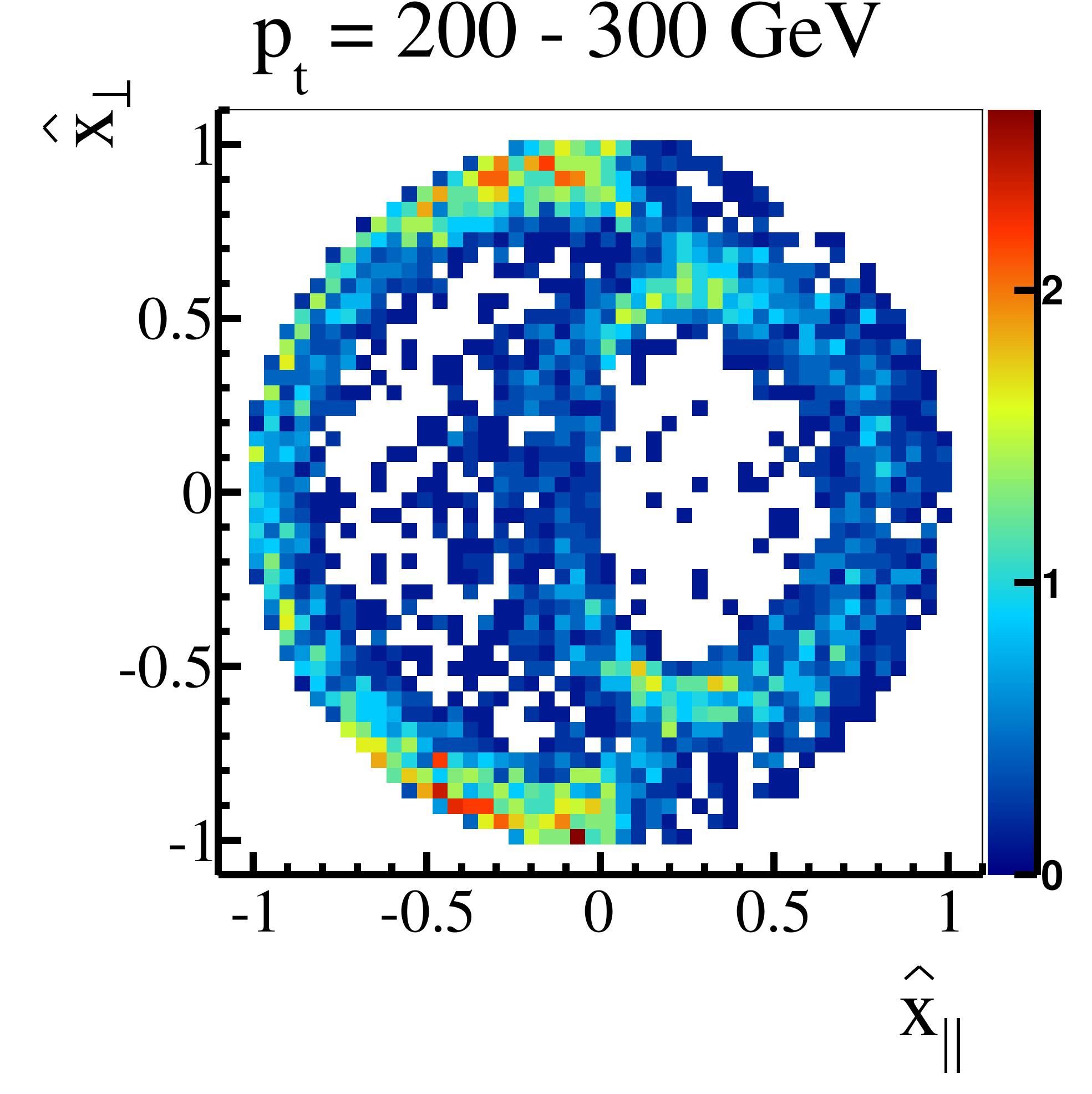}
\includegraphics[width=0.24\textwidth]{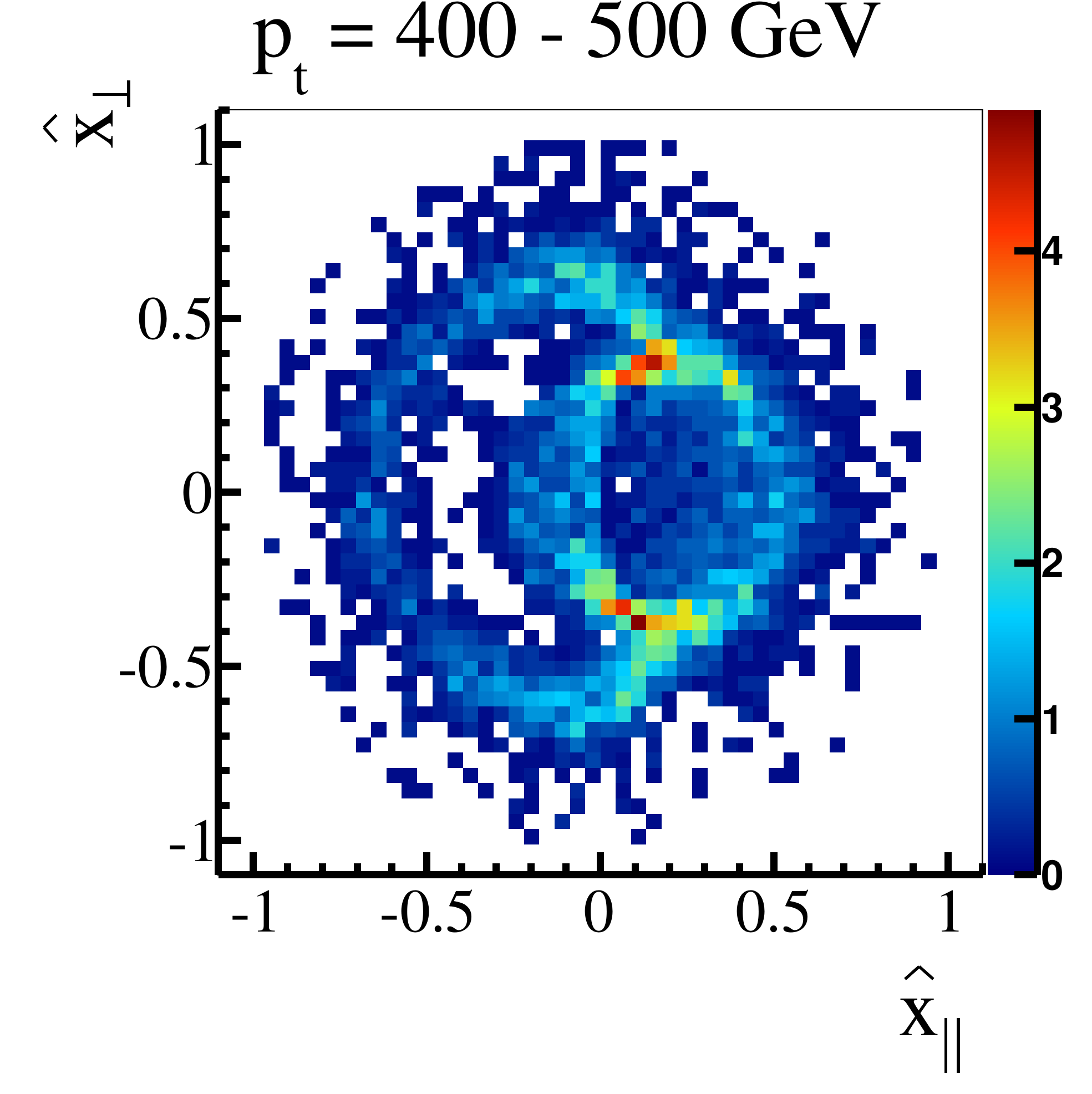}
\includegraphics[width=0.24\textwidth]{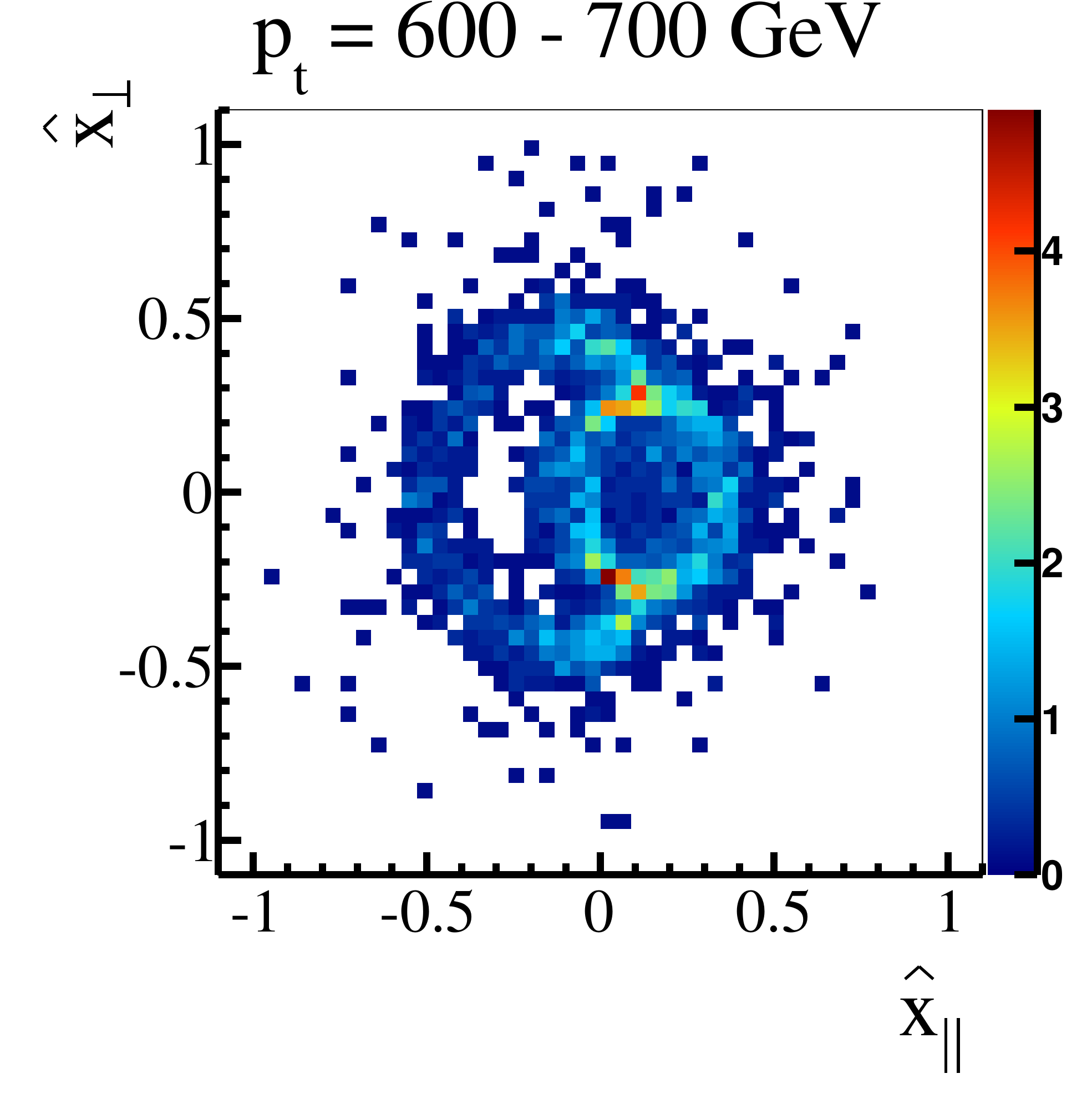}
\includegraphics[width=0.24\textwidth]{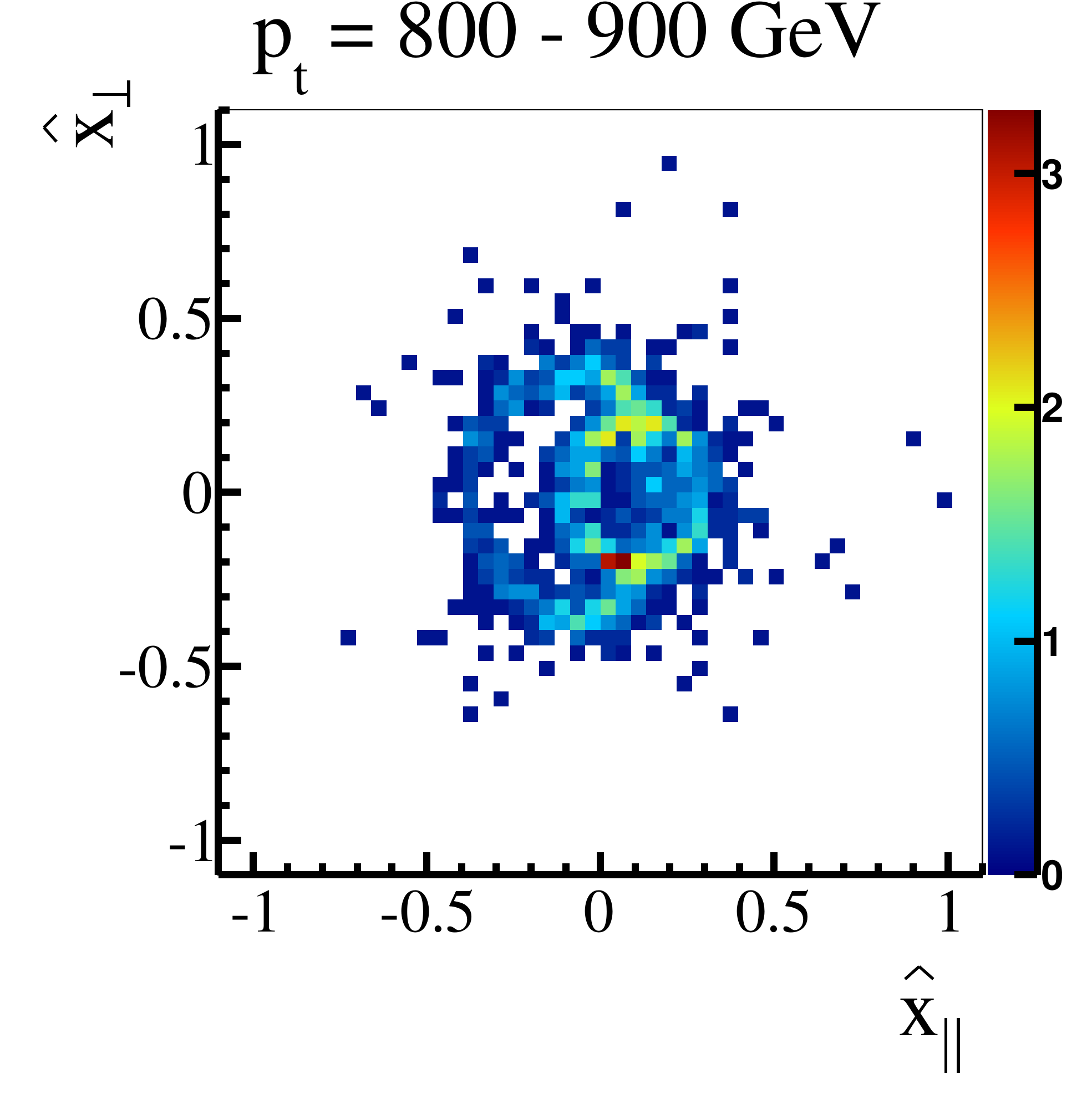}
\vspace*{-5mm}
\caption{$(\hat{x}_\parallel, \hat{x}_\perp)$ distributions as defined
  in Eq.(\ref{eq:def_xhat}) for top pair production. The four panels
  correspond to slices in the top momentum $200-300$~GeV,
  $400-500$~GeV, $600-700$~GeV, and $800-900$~GeV. All events fulfill
  $\met>150$~GeV.}
\label{fig:x2x3}
\end{figure}
%-----------------------------------------------------

In this parameterization, with $m_{b\ell}^2=(p_{\ell} + p_b)^2$, the
top mass constraint becomes an ellipse in three dimensions
\begin{alignat}{5}
  (\vec{p}^2_{b \ell} + m^2_{b \ell}) x_\parallel^2
+ (\vec{p}^2_{b \ell} + m^2_{b \ell}) x_\perp^2
+ m^2_{b \ell} x_\text{D}^2 - (m_t^2 - m_{b \ell}^2) |\vec{p}_{b \ell}| x_\text{D} 
&= (m_t^2 - m_{b \ell}^2)^2/4 \notag \\
\text{or} \qquad \qquad
\frac{x_\parallel^2}{R_{t,1}^2} 
+ \frac{x_\perp^2}{R_{t,1}^2} 
+  \frac{(x_\text{D}-\bar x_\text{D})^2}{R_{t,2}^2} 
 &= 1 \; ,
\end{alignat}
with the radii and focal point coordinate
\begin{equation}
R_{t,1} =\frac{m_t^2 -m_{b \ell}^2}{2 m_{b \ell}} \; ,
\qquad \qquad
R_{t,2} =\frac{\sqrt{|\vec{p}_{b \ell}|^2 + m_{b \ell}^2}}{m_{b \ell}} R_{t,1} \; ,
\qquad \qquad
\bar x_\text{D} =\frac{|\vec{p}_{b \ell}| }{m_{b \ell}} R_{t,1} \; .
\end{equation}
For the $W$ mass constraint it is useful to define the rotated basis vectors 
\begin{equation}
\begin{pmatrix}
y_\text{D} \cr y_\parallel
\end{pmatrix}
= 
\begin{pmatrix}
\cos\theta_{b \ell,\ell}& \sin\theta_{b \ell,\ell} \cr
-\sin\theta_{b \ell,\ell} & \cos\theta_{b \ell,\ell}
\end{pmatrix}
\begin{pmatrix}
x_\text{D} \cr x_\parallel
\end{pmatrix} \; ,
\end{equation}
using $ \cos\theta_{b \ell,\ell}=\hat{p}_{\ell} \cdot
\hat{p}^\text{D}$. In these variables the $W$ mass constraint reads
\begin{alignat}{5}
|\vec{p}_\ell|^2 y_\parallel^2
+ |\vec{p}_\ell|^2 x_\perp^2
 - m_W^2   |\vec{p}_\ell| y_\text{D}
= \frac{m_W^4}{4} \notag \\
%\text{or} \qquad \qquad
%\frac{y_\text{D}}{|\vec{p}_\ell|}
%= 
%\frac{y_\parallel^2}{m_W^2}
%+ \frac{x_\perp^2}{m_W^2}
%- \frac{m_W^2}{4|\vec{p}_\ell|^2}\notag \\
\text{or} \qquad \qquad
\frac{y_\parallel^2}{R_W^2}
+ \frac{x_\perp^2}{R_W^2}
- \frac{2 y_\text{D}}{R_W}= 1 
\end{alignat}
with the radius $R_W= m_W^2/2|\vec{p}_\ell|$.
\bigskip

We can now study the two approximations for the neutrino momentum
defined in Eq.(\ref{eq:approx}).  In the decay plane approximation
$x_\perp=0$ the two mass constraints become an ellipse and a parabola
in the $(x_\parallel, x_\text{D})$ space
\begin{equation}
\frac{x_\parallel^2}{R_{t,1}^2} 
+ \frac{(x_\text{D}-\bar x_D)^2}{R_{t,2}^2} 
=1
\qquad \qquad \text{and} \qquad \qquad 
\frac{y_\parallel^2}{R_W^2}
- \frac{2 y_\text{D}}{R_W}= 1\; .
\label{eq:decayplane}
\end{equation}
The neutrino momentum we obtain as the intersection of two lines,
where in general there exist up to four solutions. We use the solution
which gives the smallest top momentum. In the orthogonal approximation
$x_\parallel=0$ the mass conditions are the two ellipses
\begin{equation}
\frac{x_\perp^2}{R_{t,1}^2} 
+ \frac{(x_\text{D}-\bar x_D)^2}{R_{t,2}^2} 
=1
\qquad \qquad \text{and} \qquad \qquad 
\frac{x_\perp^2}{{R'_W}^2}
+ \frac{(x_\text{D} - \bar{x}_D^\prime)^2}{{R''_W}^2}
= 1 \; ,
\label{eq:orthogonal}
\end{equation} 
with the rotated radii and focal coordinate 
\begin{equation}
R'_W = \frac{R_W}{\sin\theta_{b \ell,\ell}},  \qqquad 
R''_W = \frac{R_W}{\sin^2\theta_{b \ell,\ell}},  \qqquad
\bar{x}_D^\prime=\frac{\cos\theta_{b \ell,\ell}R_{3}}{\sin^2\theta_{b \ell,\ell}} \; .
\end{equation}
We find these solutions for the orthogonal approximation numerically,
being aware of the fact that in some cases there does not actually
exist any solution. This finite efficiency we observe in our analysis
results in Table~\ref{tab:res2}. In case we find more than one
solution we keep the one which gives the smaller subtracted missing
momentum $\met - p_{T,\nu}^\text{est}$.\bigskip

%-----------------------------------------------------
\begin{figure}[t]
\includegraphics[width=0.23\textwidth]{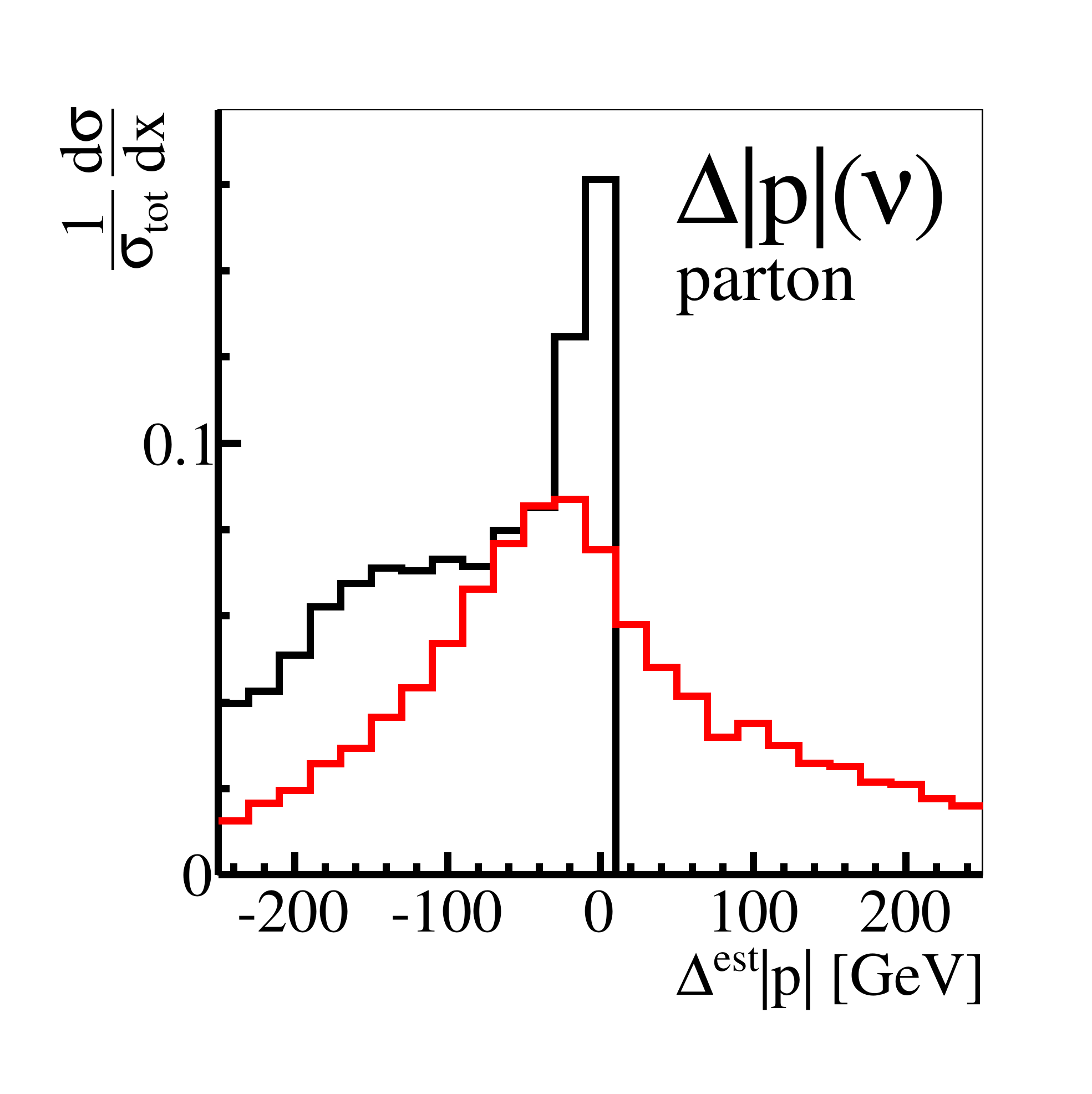}
\includegraphics[width=0.23\textwidth]{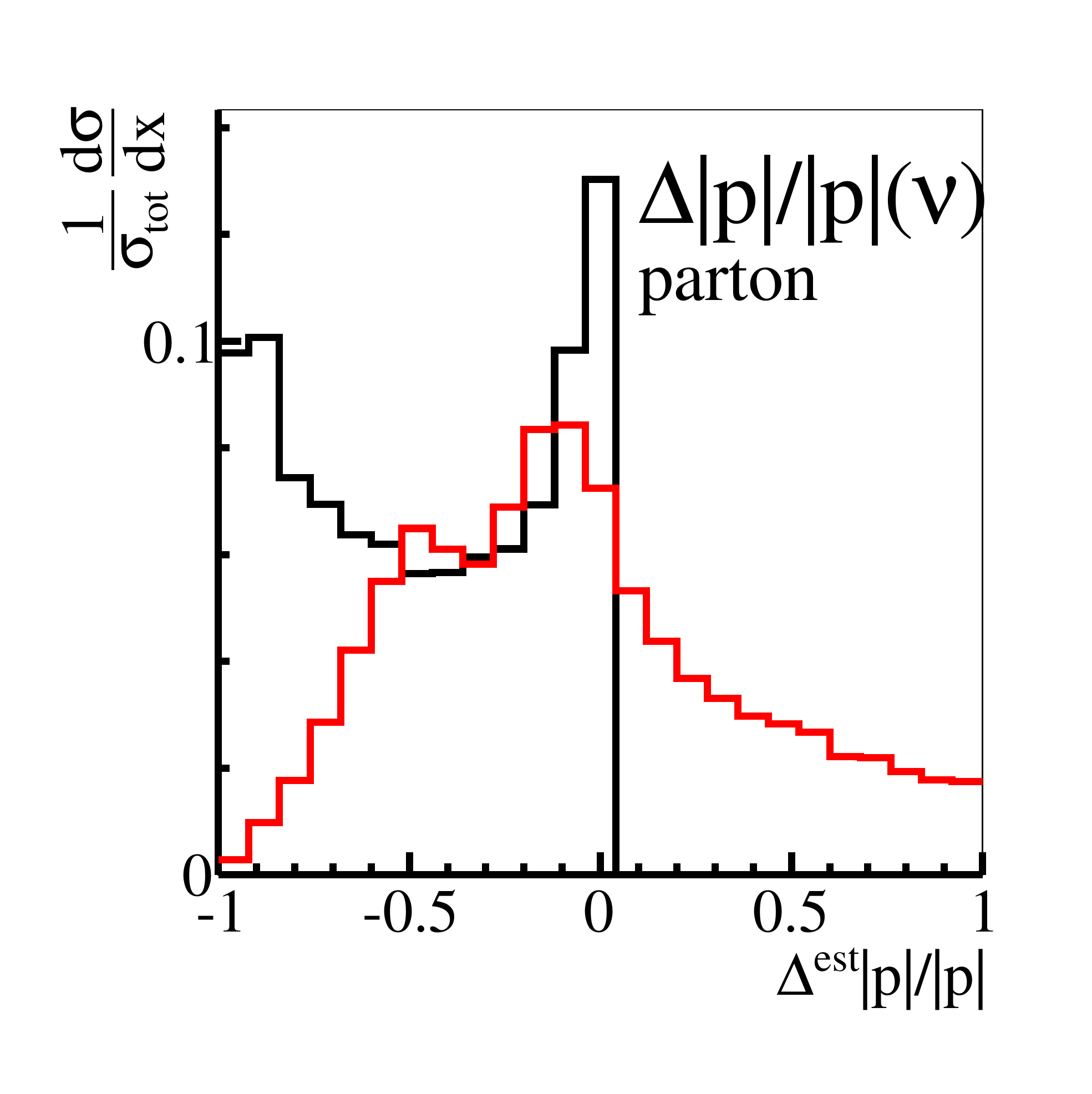}
\includegraphics[width=0.23\textwidth]{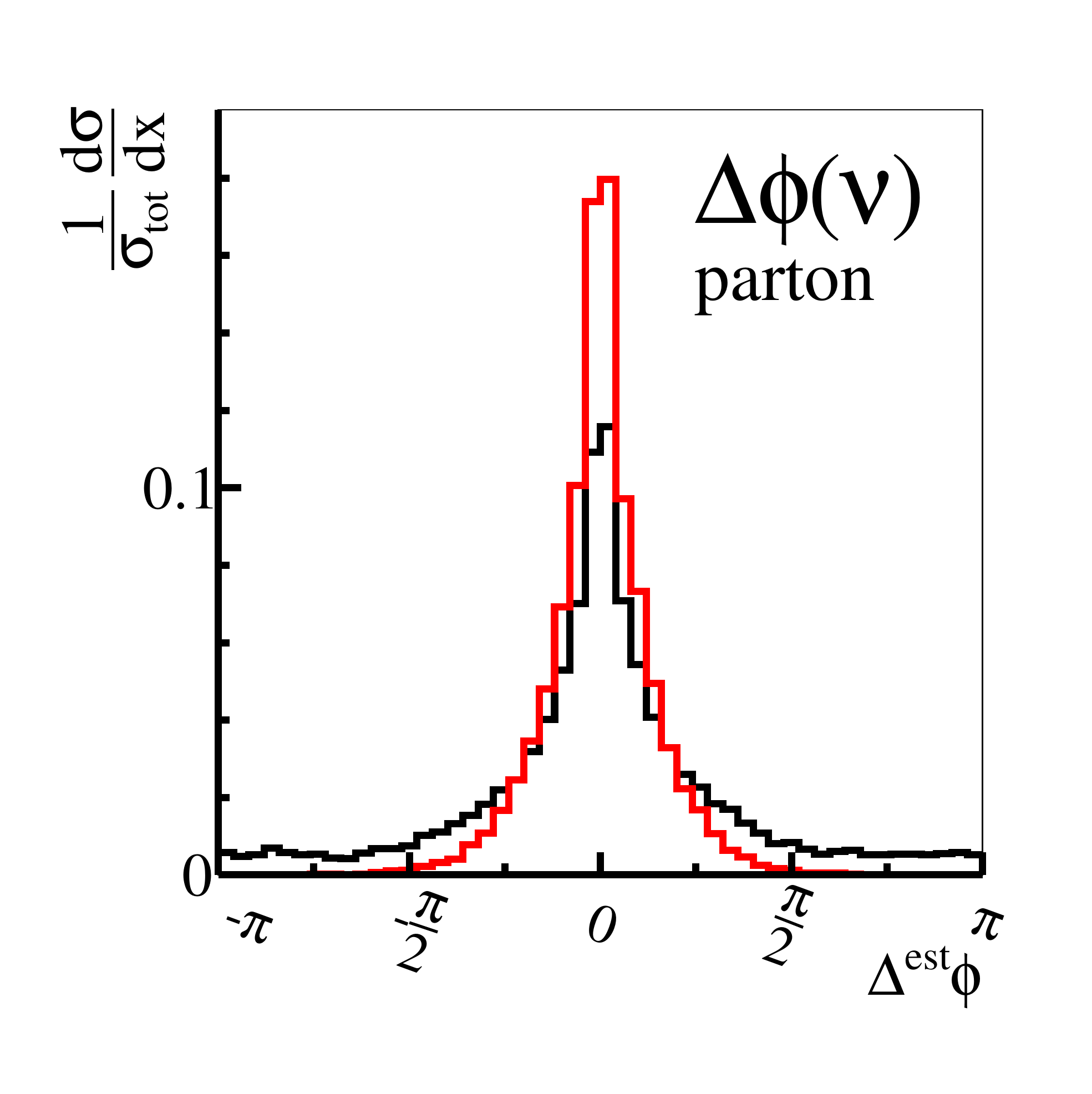}
\includegraphics[width=0.23\textwidth]{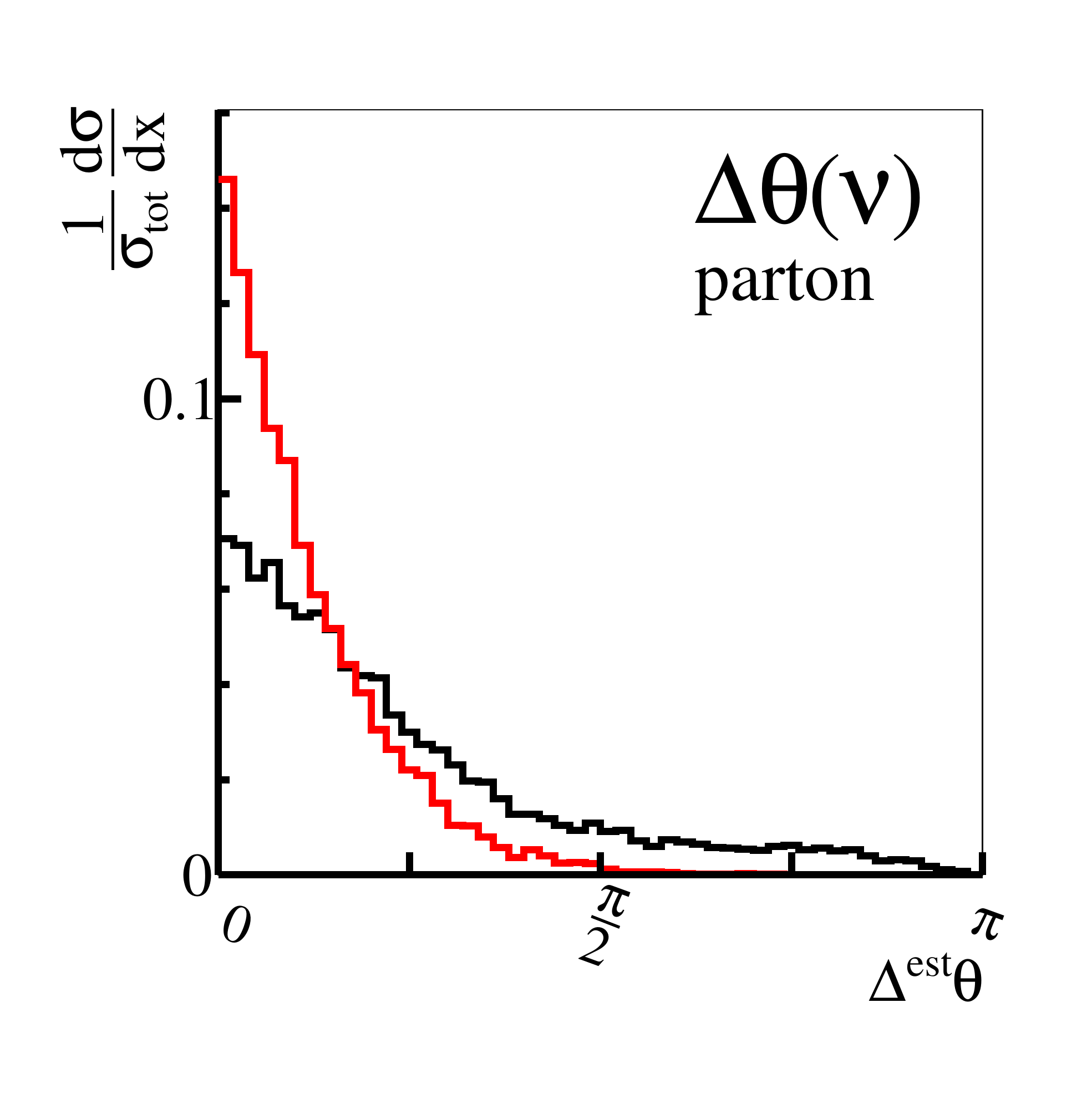}

\includegraphics[width=0.23\textwidth]{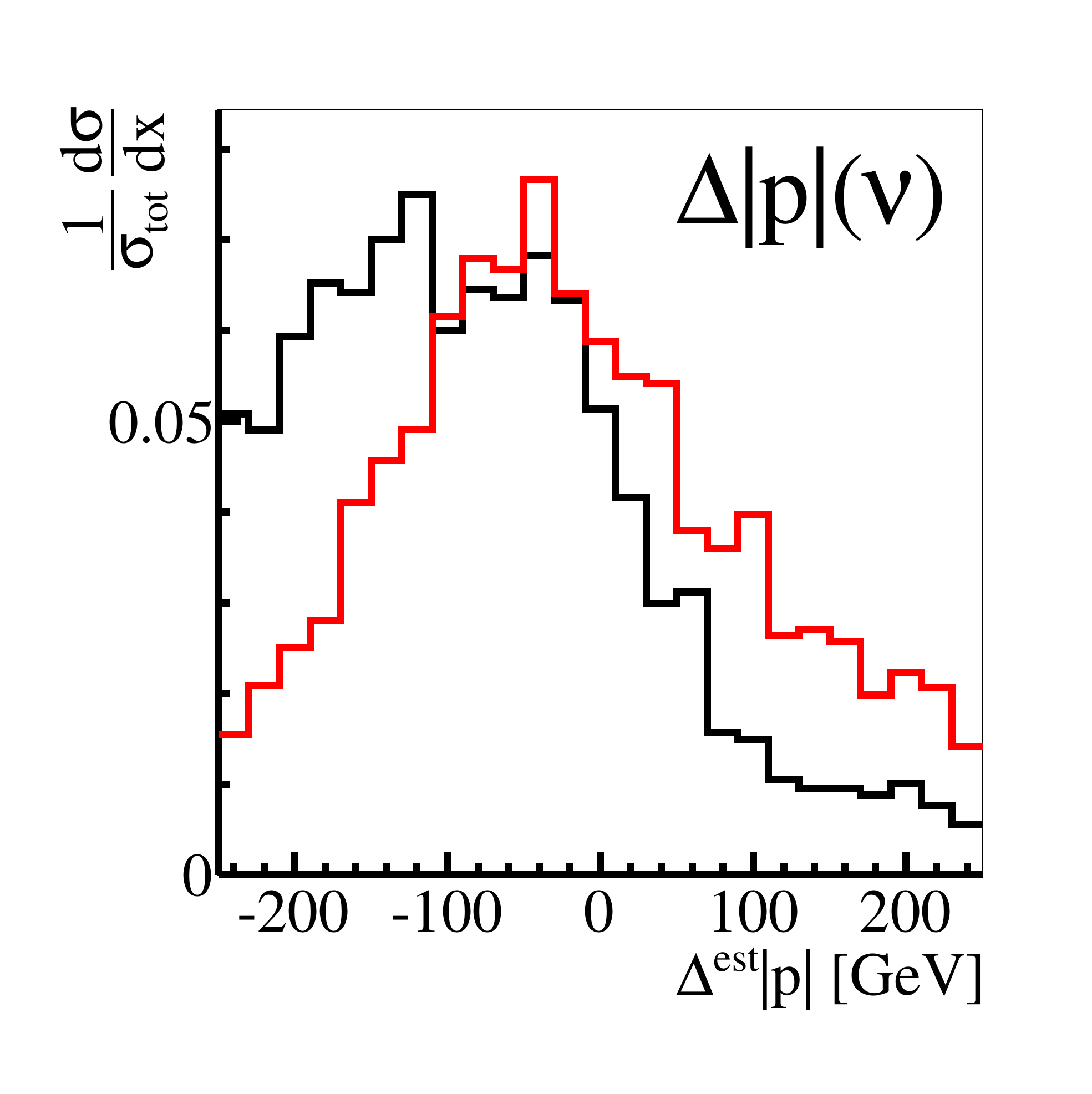}
\includegraphics[width=0.23\textwidth]{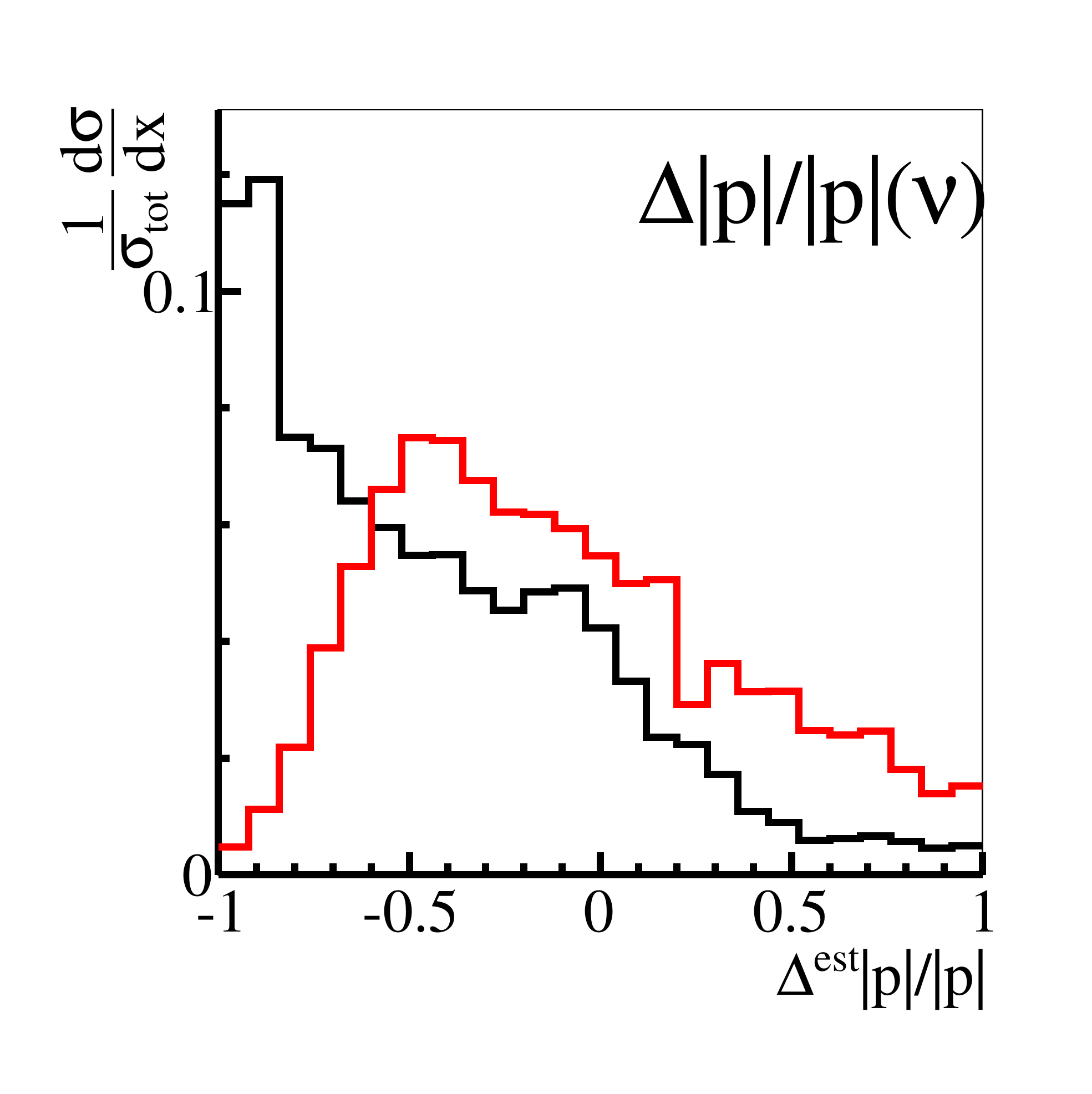}
\includegraphics[width=0.23\textwidth]{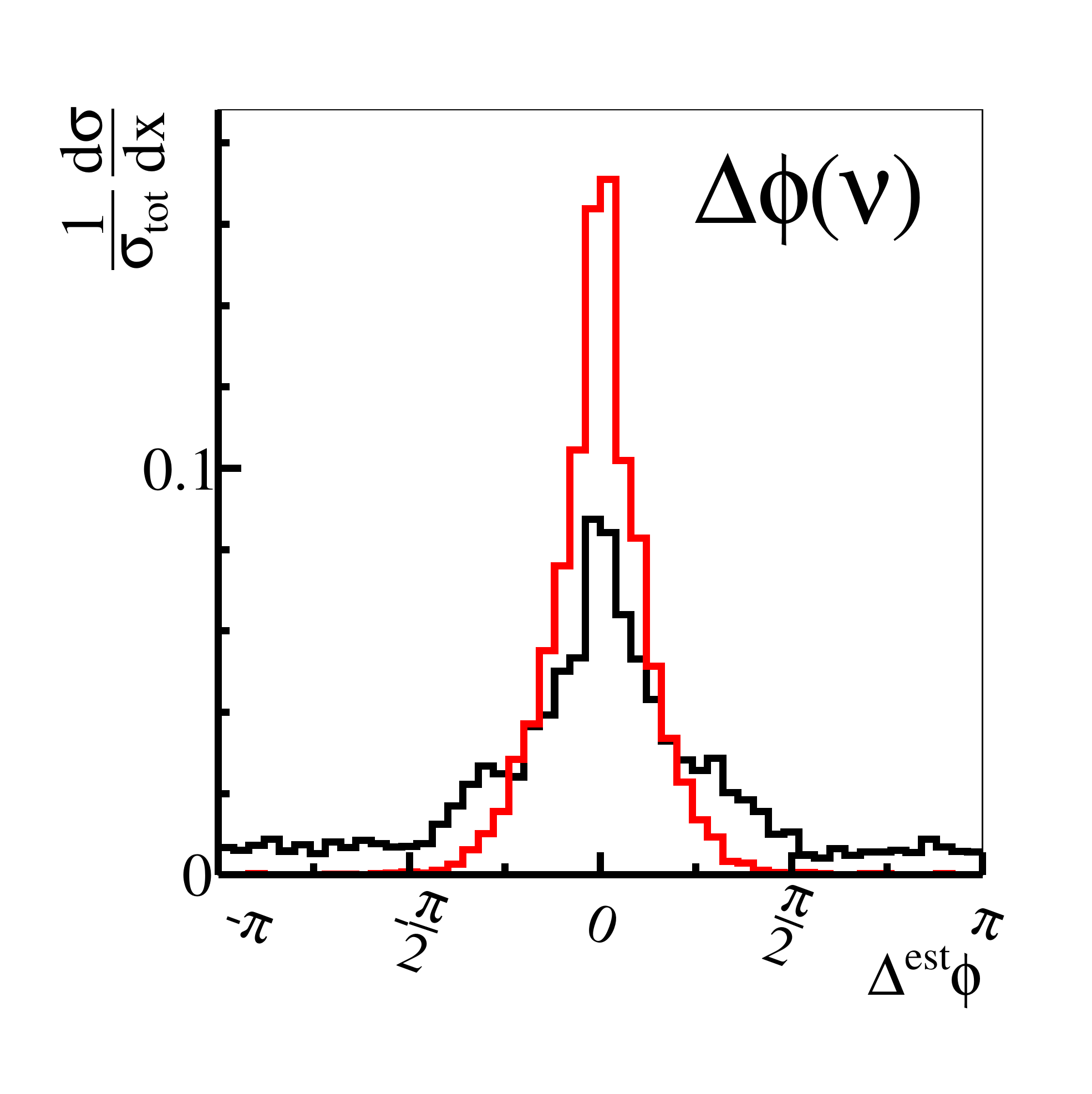}
\includegraphics[width=0.23\textwidth]{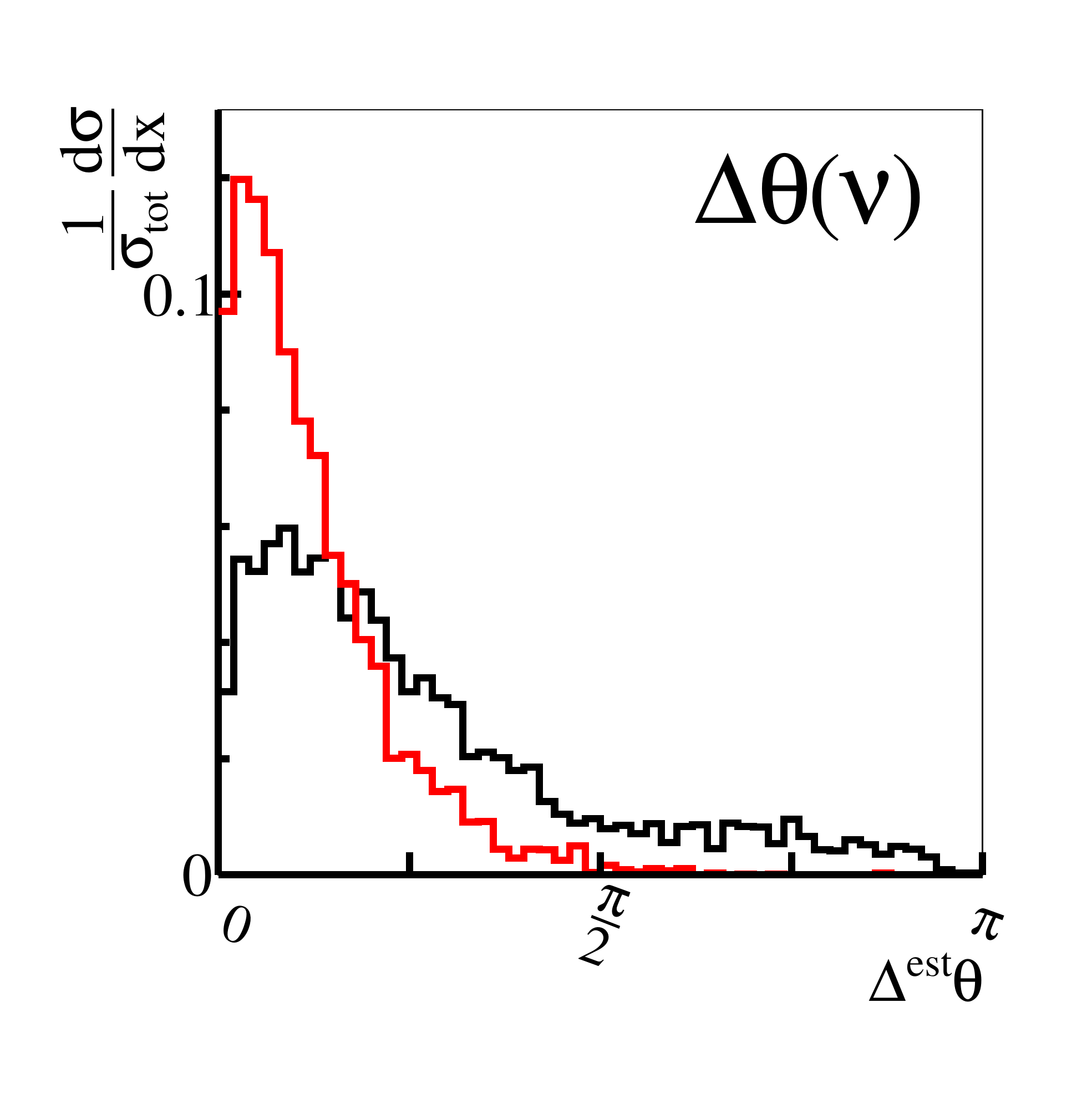}

\includegraphics[width=0.23\textwidth]{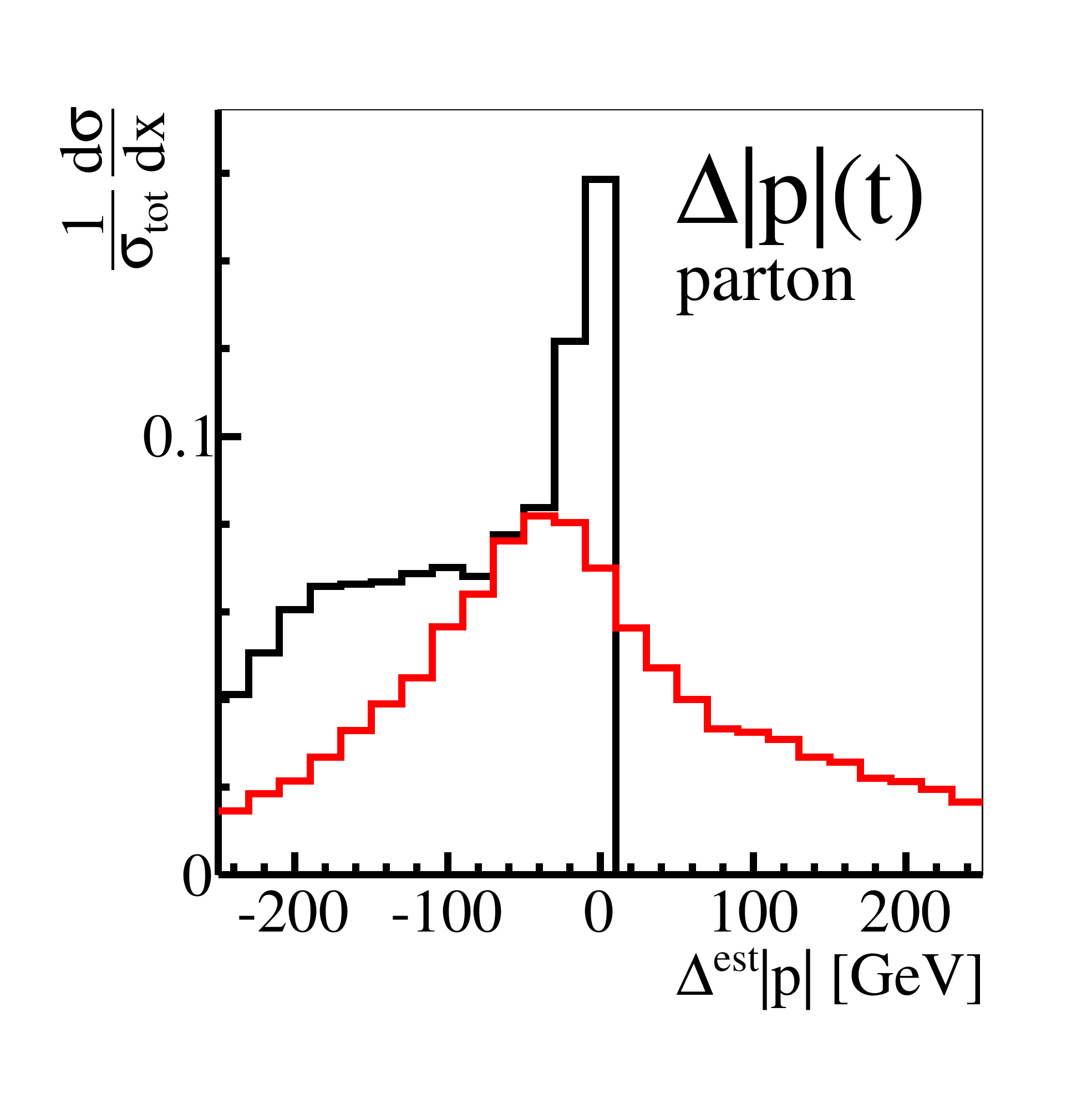}
\includegraphics[width=0.23\textwidth]{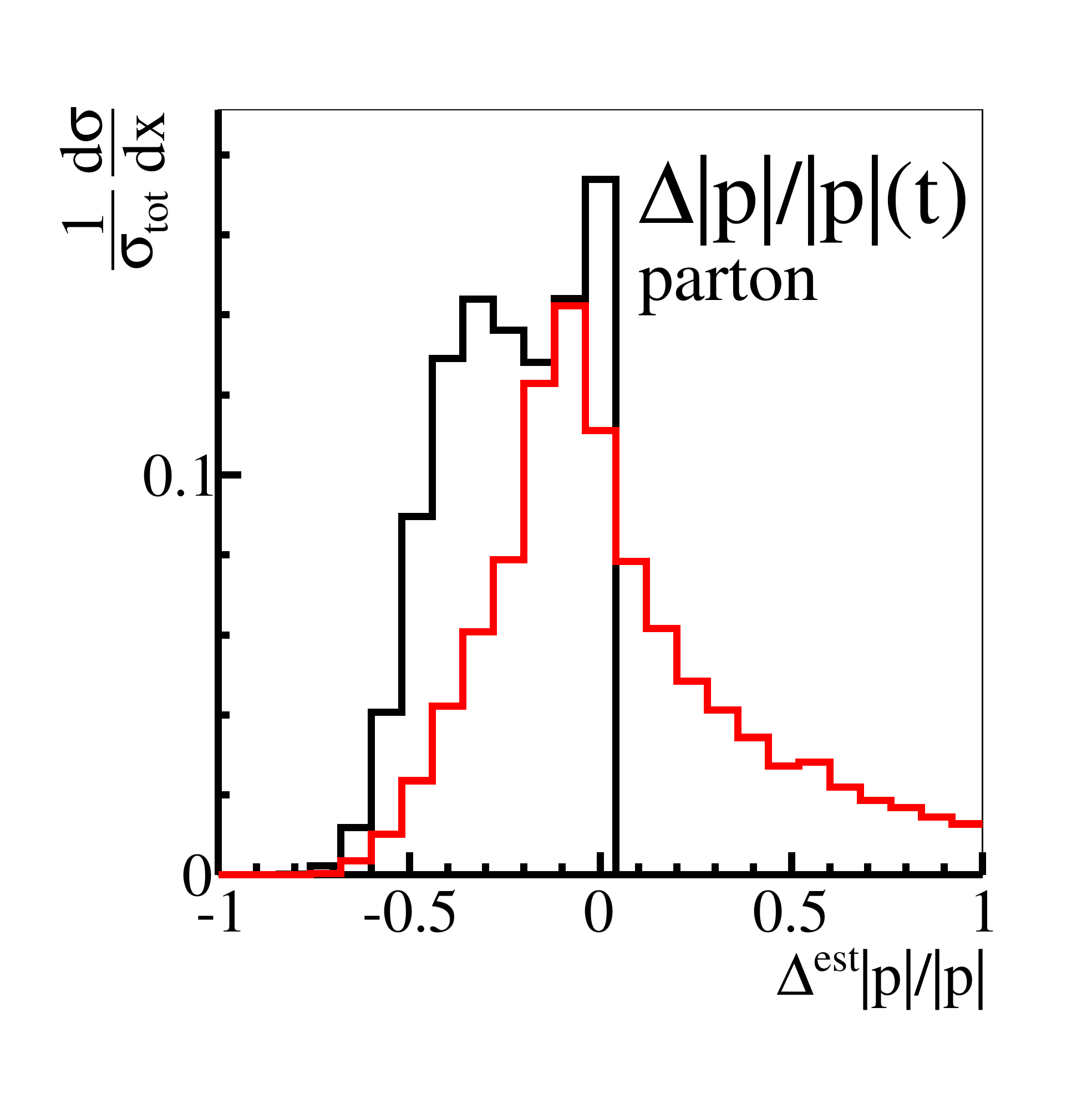}
\includegraphics[width=0.23\textwidth]{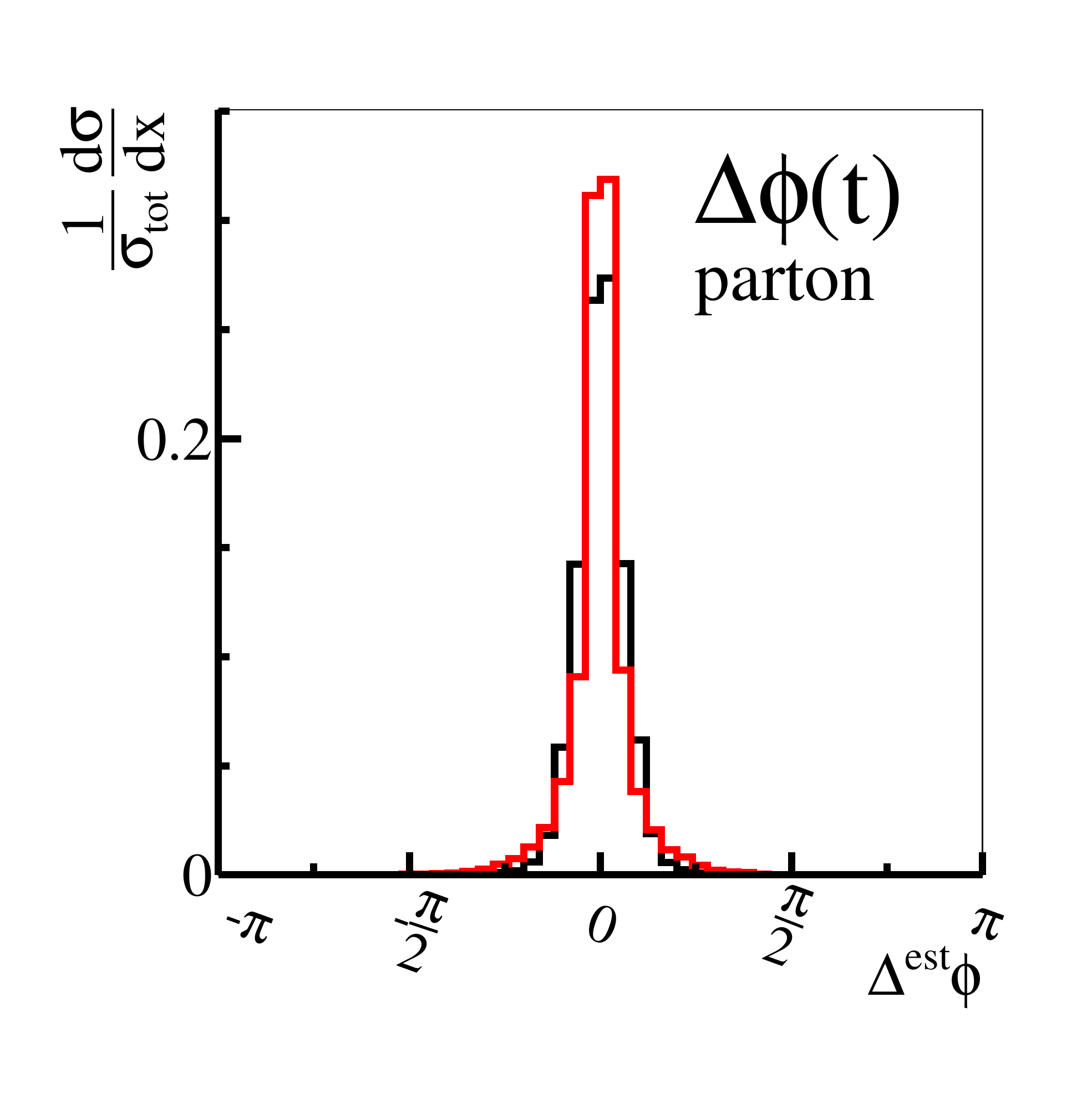}
\includegraphics[width=0.23\textwidth]{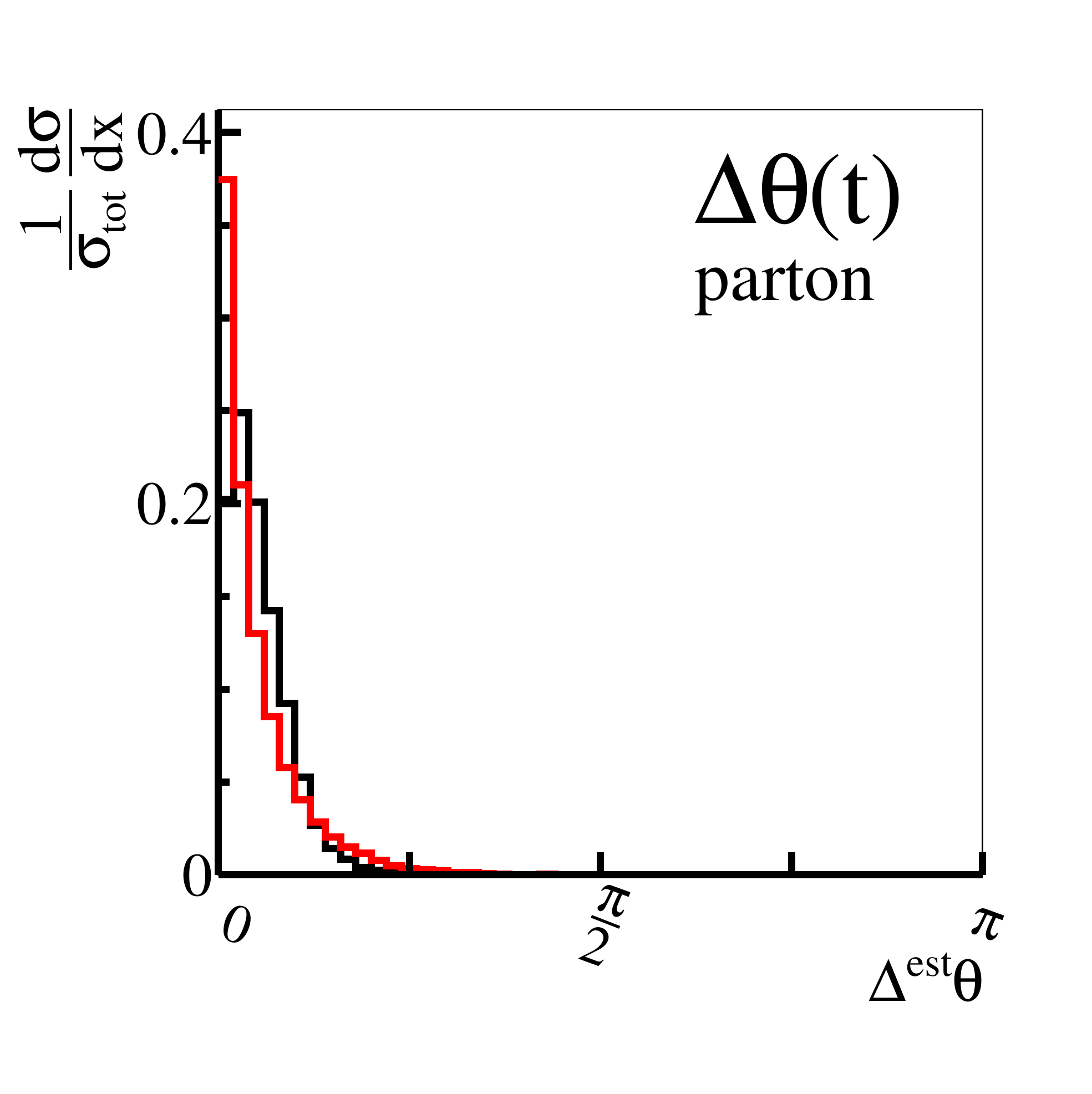}

\includegraphics[width=0.23\textwidth]{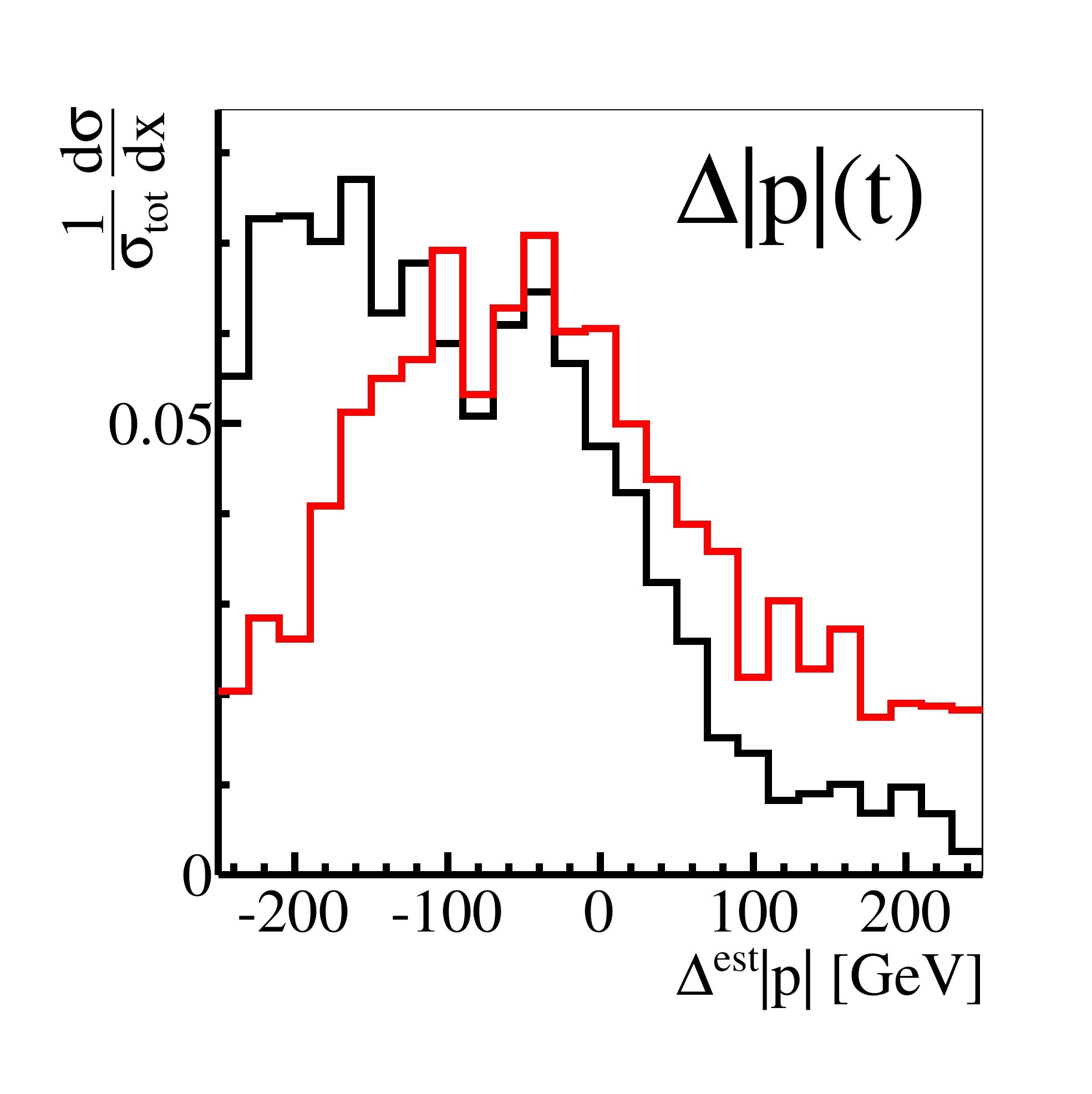}
\includegraphics[width=0.23\textwidth]{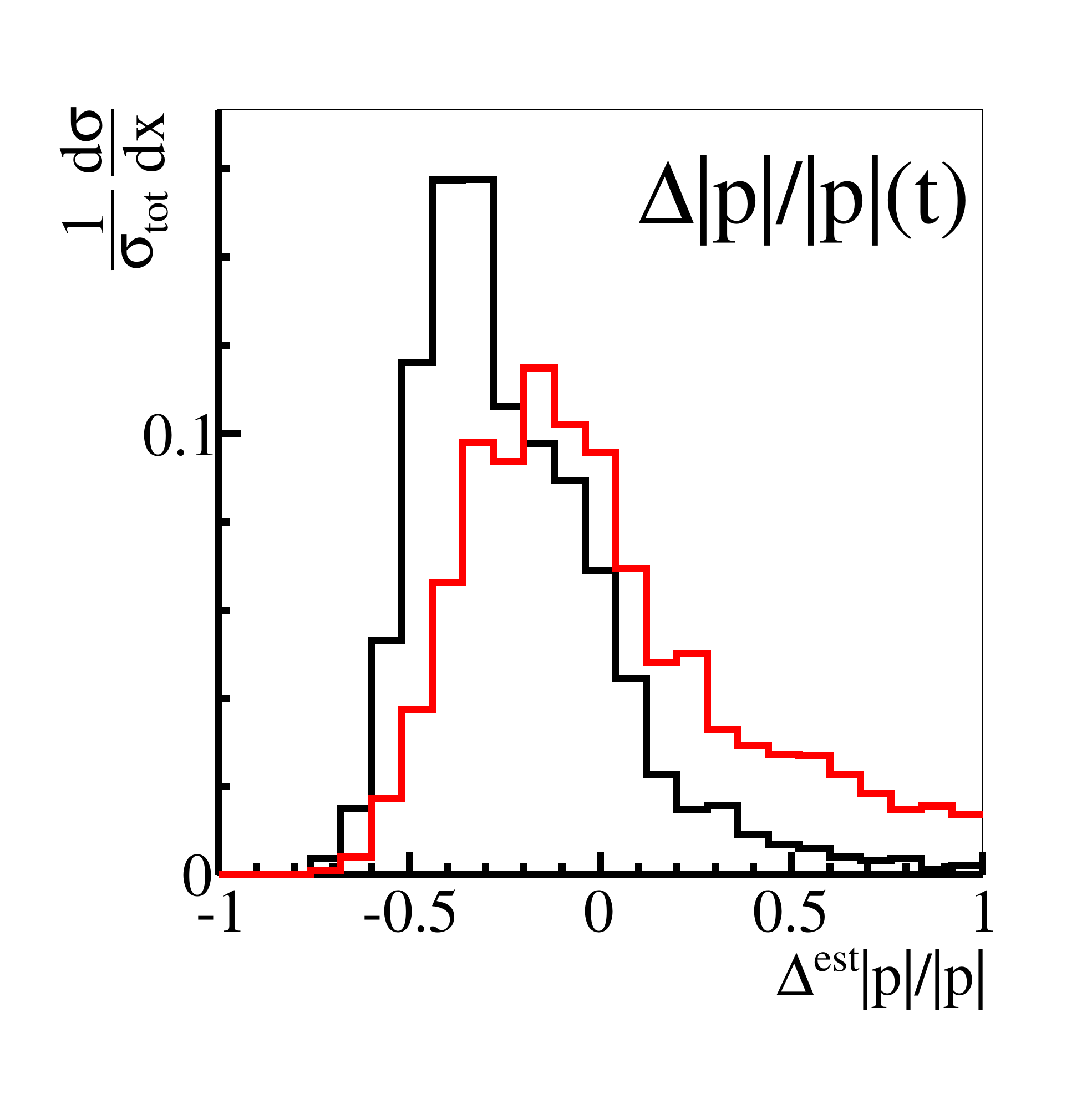}
\includegraphics[width=0.23\textwidth]{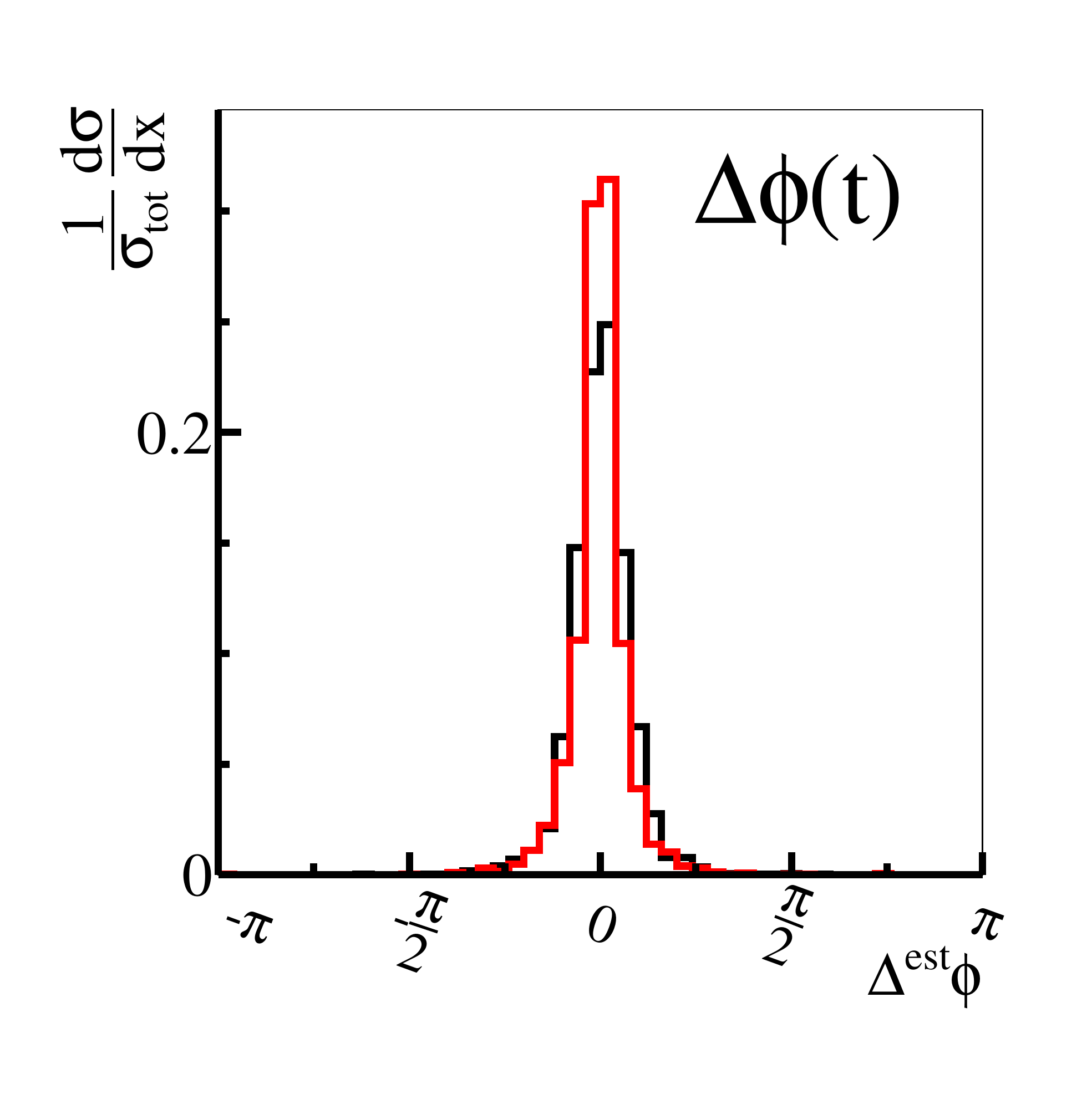}
\includegraphics[width=0.23\textwidth]{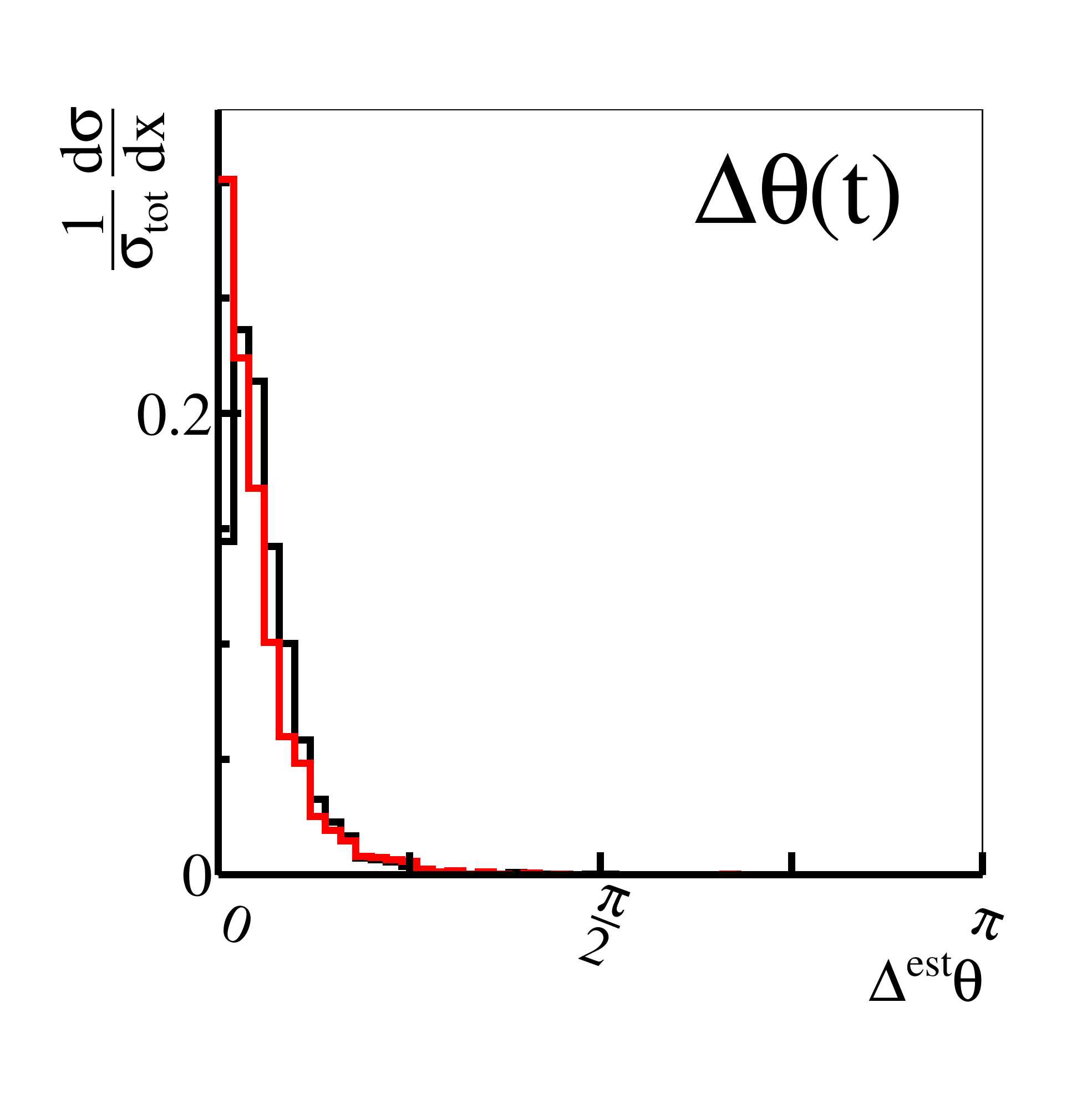}

\caption{Absolute and relative $\Delta^\text{est} |p|$,
  $\Delta^\text{est} |p|/|p^\text{true}|$, $\Delta^\text{est} \phi$
  and $\Delta^\text{est} \theta$ distributions in the decay plane
  approximation (black) and the orthogonal approximation (red). The
  upper two rows show the neutrino momentum estimate, the lower two
  rows the top momentum estimate.  The first of the two rows is at
  parton level, the second row at hadron level. Only semileptonic top
  pair events are included. At hadron level we only require
  $\met>150$~GeV while at parton level all events also fulfill
  $p_{T,\ell} > 20$~GeV and $p_{T,b} > 25$~GeV.}
\label{fig:dphi}
\end{figure}
%-----------------------------------------------------

In Figure~\ref{fig:dphi} we show several measures describing the
difference between the estimated leptonic top and neutrino momenta and
the respective true momenta at parton level:
\begin{alignat}{5}
\Delta^\text{est} |p_{t, \nu}| 
& = |p_{t, \nu}^\text{est}| - |p_{t, \nu}^\text{true}| 
\notag \\
\Delta^\text{est} \phi_{t, \nu} 
&= \phi_{t, \nu}^\text{est} - \phi_{t, \nu}^\text{true}
\notag \\
\Delta^\text{est} \theta_{t, \nu} 
&= \theta_{t, \nu}^\text{est} - \theta_{t, \nu}^\text{true}
\end{alignat}
At hadron level the orthogonal approximation shown in red performs
consistently better than the decay plane approximation to reconstruct
the direction of the neutrino and the top momenta. We also see that
the azimuthal direction of the top quark can be more precisely
reconstructed than the neutrino's azimuthal direction. This occurs
because compared to the bottom and lepton reconstruction the neutrino
is the weak spot of the top reconstruction.  This is why in our
analysis described in Section~\ref{sec:dir} we use the top azimuthal
direction estimated in the orthogonal approximation. Only considering
tops with $p_{T,t} > 300$~GeV only slightly improves this picture.

If we do not apply the condition $\met > 150$~GeV effectively on the
neutrino momentum the corresponding analysis to Figure~\ref{fig:dphi}
paints a different picture: the decay plane approximation achieves the
better resolution at parton level, and this features is largely the
same at hadron level. This is the reason why our $M_{T,2}$ endpoint
analysis is based on a mass reconstruction in the decay approximation
(after the event selection in the orthogonal approximation).

%%%%%%%%%%%%%%%%%%%%%%%%%%%%%%%%%%%%%%%%%%%%%
\section{Jacobian in the decay plane approximation}
\label{app:approx}

The relative position of the top decay products is completely
determined by the vector of variables
$X=(\gamma_t,\cos\theta_b,\cos\theta,\phi)$. Here, $\gamma_t$ is the
boost factor of the top in the laboratory frame
$p_t^\text{lab}=\beta_t \gamma_t m_t$, $\theta_b$ is the $t \to bW$
decay angle with respect to the top direction, and $(\theta, \phi)$
are the $W$ decay angles in the $W$ rest frame. In terms of these
variables, the three momenta of the decay products in the top rest
frame are given by
\begin{equation}
p_b=
\begin{pmatrix}
E_b\\
0 \\
0\\ 
|\vec{p}_b|
\end{pmatrix},
\qqquad
p_\ell =
\frac{m_W}{2} 
\begin{pmatrix}
\gamma - \gamma \beta \cos\theta \\
\sin\theta\cos\phi \\
\sin\theta\sin\phi\\ 
\gamma \cos\theta -\gamma \beta
\end{pmatrix},
\qqquad 
p_\nu =
\frac{m_W}{2} 
\begin{pmatrix}
\gamma + \gamma \beta \cos\theta \\
- \sin\theta\cos\phi \\
- \sin\theta\sin\phi\\ 
- \gamma \cos\theta -\gamma \beta
\end{pmatrix} \; .
\end{equation}
$\beta$ and $\gamma$ are defined by the $W$ and $b$ momenta
in the top rest frame
\begin{equation}
\beta\gamma m_W =
|\vec{p}_W| = 
|\vec{p}_b| = 
\frac{m_t}{2} 
\sqrt{1-\left(\frac{m_W+m_b}{m_t}\right)^2}
\sqrt{1-\left(\frac{m_W-m_b}{m_t}\right)^2} \; ,
\end{equation}
and $E_b=\sqrt{ \vec{p}_b^2 + m_b^2}$. For an unpolarized top the top
decay phase space distribution in terms of the variable vector $X$ can
be written as~\cite{Kane:1991bg}
\begin{equation}
 P(X) = P_t(\gamma_t) \; P_W(\cos\theta) \; ,
\end{equation}
where $P_t \sim \exp(- \text{const} \times \gamma_t)/\gamma_t$ is the
distribution for the top momentum and $P_W$ is the distribution for
the $W$ decay angle $\theta$.\bigskip

The problem of $P(X)$ is that it does not correspond to our
measurements listed in Eq.(\ref{eq:measurements}). Linking the
observables $X$ to our energy measurements in the laboratory frame
returns a Jacobian which dominates the distribution of the top decay
products in our analysis.  After boosting the lepton--bottom system by
$\gamma_t$ in the direction of $\theta_b$ the lepton and $b$ energies
and their invariant mass from Eq.(\ref{eq:measurements}) become
\begin{alignat}{5}
E_b^\text{lab}&= \gamma_t E_b + \gamma_t \beta_t |\vec{p}_b| \cos\theta_b, \cr
E_{\ell}^\text{lab}&=\frac{m_W}{2} 
\Big[ \gamma_t \gamma (1-\beta \cos \theta) 
- \gamma_t \beta_t \sin\theta_b \sin\theta\sin\phi
+ \gamma_t \beta_t \gamma \cos\theta_b (\cos \theta-\beta ) \Big],
\cr
m_{b\ell}^2 &=
\frac{m_t^2 - m_W^2 + m_b^2}{2} - m_t |\vec{p}_b| \cos\theta \; , 
\label{eq:observable}
\end{alignat}
These observables we can express in terms of dimensionless vector
$Y=(\gamma_t,\tilde{E}_b,\tilde{E}_{\ell},\tilde{m}_{b\ell}^2)$
including
\begin{equation}
\tilde{E}_b= \frac{E_b^\text{lab}}{|\vec{p}_b|} , \qqquad
\tilde{E}_{\ell}= 2 \frac{E_{\ell}}{m_W} , \qqquad 
\tilde{m}_{b\ell}^2= \frac{m_{b\ell}^2}{m_t |\vec{p}_b|} \; .
\end{equation}
Using the relation $P(X)dX = P_Y(Y)dY$, the distribution $P_Y(Y)$
after a variable transformation can be calculated from the identity
\begin{alignat}{5}
P_Y(Y) =& 
\det \left|\frac{\partial Y}{\partial X}\right|^{-1} \; 
P_t(\gamma_t) \; 
P_W\left(\frac{m_t^2 - m_W^2 + m_b^2}{2 m_t |\vec{p}_b|} - {\tilde m}_{b\ell}^2 \right) 
\notag \\
\text{with} \quad 
& \det \left|\frac{\partial Y}{\partial X}\right|=
\gamma_t^2  \, \beta_t^2 \, \sin\theta_b \, \sin\theta \, |\cos\phi| 
= \gamma_t \beta_t 
\sqrt{A\gamma_t^2 + 2B\gamma_t + C} \; ,
\label{eq:prob}
\end{alignat}
in terms of $\theta$ given by the last line of
Eq.(\ref{eq:observable}) and 
\begin{alignat}{5}
A&=- \sin^2 \theta \, \left( \frac{E_b^2}{|\vec{p}_b|^2} -1 \right)
   - \gamma^2 \sin^4 \theta \, (1+\beta)^2 \notag \\
B&= \sin^2 \theta \, \frac{{\tilde E}_b E_b}{|\vec{p}_b|}
   - \gamma \sin^2 \theta \, (1+\beta) \, 
      [{\tilde E}_{\ell} - \gamma(\cos \theta - \beta) {\tilde E}_b] \notag \\
C&=- \sin^2 \theta \, (1+{\tilde E}_b^2)
   - [{\tilde E}_{\ell} - \gamma( \cos \theta - \beta) {\tilde E}_b]^2,
\end{alignat}
Once we measure the $Y$ components $({\tilde E}_b, {\tilde E}_{\ell},
{\tilde m}_{b\ell})$ we can estimate the top momentum according to
this probability $P_Y(Y)$. The poles in the Jacobian $\det |\partial
Y/\partial X|=0$ give us a
pronounced Jacobian peak and constrain the allowed range of the top
momentum.  Following Eq.(\ref{eq:prob}) the peak position implies
$\cos\phi=0$, which corresponds to the kinematic configuration where
$\vec{p}_b$, $\vec{p}_{\ell}$ and $\vec{p}_\nu$ are coplanar.

Moreover, the pole condition is a quadratic equation for $\gamma_t$
which has two solutions. Because $P_t(\gamma_t)$ is steeply falling
for increasing $\gamma_t$ the most likely value for $\gamma_t$ is the
smaller one.  To estimate the top quark momentum we therefore choose
the solution which yields the smaller momentum in the decay plane
approximation ($x_\perp=0$).

%%%%%%%%%%%%%%%%%%%%%%%%%%%%%%%%%%%%%%%%%%%%%%%%%%%%%%%%%%%%%%%%%%%%%%%%%%%%%%%%

\end{document}